\documentclass[12pt]{iopart}
\usepackage[pdftex]{graphicx}
\expandafter\let\csname equation*\endcsname\relax
\expandafter\let\csname endequation*\endcsname\relax
\usepackage{amssymb}
\usepackage{amsmath}
\usepackage{multirow}
\usepackage{url}
\usepackage{color}
\usepackage{textcomp}
\usepackage{hyperref}
\usepackage{orcidlink}
\usepackage{acronym}

\newcommand{\secref}[1]{section~\ref{#1}}
\newcommand{\figref}[1]{figure~\ref{#1}}
\newcommand{\tabref}[1]{table~\ref{#1}}

\newcommand{\Figref}[1]{Figure~\ref{#1}}

\newcommand{\coord}{coordinate}
\newcommand{\Coord}{Coordinate}

\newcommand{\nint}[1]{\left\lfloor#1\right\rceil}

\newcommand{\un}[1]{\text{\,#1}}

\newcommand{\ev}[1]{E\left[#1\right]}
\newcommand{\Tsft}{T_{\text{sft}}}
\newcommand{\Tmax}{T_{\text{max}}}
\newcommand{\Tobs}{T_{\text{obs}}}
\newcommand{\Trun}{\Tobs}
\newcommand{\muobs}{\mu_{\text{obs}}} \newcommand{\sigobs}{\sigma_{\text{obs}}} 
\newcommand{\Tasc}{t_{\text{asc}}}
\newcommand{\TascThen}{\Tasc} \newcommand{\TascNow}{\Tasc'} \newcommand{\TascNowMin}{t'_{\text{asc,min}}}
\newcommand{\TascNowMax}{t'_{\text{asc,max}}}

\newcommand{\Porb}{P_{\text{orb}}}
\newcommand{\PorbShear}{\tilde{P}}
\newcommand{\dPdTShear}{\left(\frac{\partial\Porb}{\partial\TascNow}\right)_{\PorbShear}}
\newcommand{\norb}{n_{\text{orb}}}
\newcommand{\sigPorb}{\sigma_{\Porb}}
\newcommand{\sigPorbShear}{\sigma_{\PorbShear}}
\newcommand{\sigTasc}{\sigma_{\Tasc}}
\newcommand{\sigTascThen}{\sigma_{\TascThen}} \newcommand{\sigTascNow}{\sigma_{\TascNow}} \newcommand{\ap}{a_p}
\newcommand{\gff}{g_{f_0f_0}}
\newcommand{\gaa}{g_{\ap\ap}}

\newcommand{\gNowTT}{g'_{\TascNow\TascNow}}
\newcommand{\gNowTP}{g'_{\TascNow\Porb}}
\newcommand{\gNowPT}{g'_{\Porb\TascNow}}
\newcommand{\gNowPP}{g'_{\Porb\Porb}}
\newcommand{\gShearTT}{\tilde{g}_{\TascNow\TascNow}}
\newcommand{\gShearTP}{\tilde{g}_{\TascNow\PorbShear}}

\newcommand{\gShearPP}{\tilde{g}_{\PorbShear\PorbShear}}
\newcommand{\ginvShearPP}{\tilde{g}^{\PorbShear\PorbShear}}
\newcommand{\BestTascThen}{t_{\text{asc},0}}
\newcommand{\BestTascNow}{t'_{\text{asc},0}}
\newcommand{\BestPorb}{P_0}
\newcommand{\tdet}{t}

\newcommand{\Tbar}{\overline{\tdet}}

\newcommand{\Tdiff}{\Delta\tdet}

\newcommand{\mumax}{\mu_{\text{max}}}
\newcommand{\Zn}{\mathbb{Z}^n}
\newcommand{\Ans}{A_n^*}
\newcommand{\LatticeTiling}{\texttt{LatticeTiling}}

\DeclareMathOperator{\sinc}{sinc}

\newcommand{\dcc}{LIGO-P2000502-v3}

\numberwithin{equation}{section}
\acrodef{ScoX1}[Sco~X-1]{Scorpius~X-1}
\acrodef{GW}[GW]{gravitational wave}
\acrodef{SNR}[SNR]{signal-to-noise ratio}

 \newcommand{\CostChopHand}{$1.434\times 10^{18}$}
\newcommand{\NumChopHand}{$1.060\times 10^{12}$}
\newcommand{\CostCubeEllip}{$2.085\times 10^{18}$}
\newcommand{\NumCubeEllip}{$1.682\times 10^{12}$}
\newcommand{\CostPropEllip}{$7.439\times 10^{17}$}
\newcommand{\NumPropEllip}{$6.006\times 10^{11}$}
\newcommand{\CostShearFour}{$7.352\times 10^{17}$}
\newcommand{\NumShearFour}{$5.931\times 10^{11}$}
\newcommand{\CostShearDiag}{$7.316\times 10^{17}$}
\newcommand{\NumShearDiag}{$5.918\times 10^{11}$}
\newcommand{\CostShearThree}{$4.928\times 10^{17}$}
\newcommand{\NumShearThree}{$3.867\times 10^{11}$}
\newcommand{\CostReallocThree}{$4.483\times 10^{17}$}
\newcommand{\NumReallocThree}{$3.431\times 10^{11}$}

\newcommand{\muObsDayTime}{2019--Sep--27 03:12:23\,UTC}

\newcommand{\muObsGPS}{1253589161}
\newcommand{\bestNorb}{4104}
\newcommand{\bestTascGPS}{1253586547}
\newcommand{\bestTascDayTime}{2019--Sep--27 02:28:49\,UTC}

\newcommand{\shearBestNorb}{4108}
\newcommand{\shearBestTascGPS}{1253858643}
\newcommand{\shearBestTascDayTime}{2019--Sep--30 06:03:45\,UTC}

\newcommand{\WangPorbSec}{68023.86}
\newcommand{\WangdPorbSec}{0.043}
\newcommand{\WangTascGPS}{974416624}
\newcommand{\WangTascDayTime}{2010--Nov--21 23:16:49\,UTC}

\newcommand{\WangdTascGPS}{50}
\newcommand{\WangAsiniMin}{1.44}
\newcommand{\WangAsiniMax}{3.25}
\newcommand{\OiiiaStartGPS}{1238112018}
\newcommand{\OiiiaStartDayTime}{2019--Apr--01 00:00:00\,UTC}

\newcommand{\OiiiaEndGPS}{1253923218}
\newcommand{\OiiiaEndDayTime}{2019--Oct--01 00:00:00\,UTC}

\newcommand{\OiiibStartGPS}{1256655618}
\newcommand{\OiiibStartDayTime}{2019--Nov--01 15:00:00\,UTC}

\newcommand{\OiiibEndGPS}{1269363618}
\newcommand{\OiiibEndDayTime}{2020--Mar--27 17:00:00\,UTC}

\newcommand{\pctonesig}{39.3}
\newcommand{\pcttwosig}{86.5}
\newcommand{\pcttresig}{98.9}
\newcommand{\pcttrepttresig}{99.6}
 
\begin{document}
\title[Template Lattices for Sco X-1 Cross-Correlation]
{Template Lattices for a Cross-Correlation Search
  for Gravitational Waves from Scorpius X-1}
\author{Katelyn J.\ Wagner\,\orcidlink{0000-0002-7255-4251}$^{1}$,
  John T.\ Whelan\,\orcidlink{0000-0001-5710-6576}$^{2,3}$,
  Jared K.\ Wofford\,\orcidlink{0000-0002-4301-2859}$^{1,3}$,
  and
  Karl Wette\,\orcidlink{0000-0002-4394-7179}$^{4,5}$
}
\ead{kjw4822@rit.edu}
\ead{john.whelan@astro.rit.edu}
\address{$^{1}$Center for Computational Relativity and Gravitation
  and School of Physics and Astronomy, Rochester Institute of Technology,
  84~Lomb Memorial Drive, Rochester, NY 14623, United States of America}
\address{$^{2}$Center for Computational Relativity and Gravitation
  and School of Mathematical Sciences, Rochester Institute of Technology,
  85~Lomb Memorial Drive, Rochester, NY 14623, United States of America}
\address{$^{3}$Institute for Theoretical Physics, Goethe University
  Frankfurt, Max-von-Laue Str.~1, D-60438 Frankfurt am Main, Germany}
\address{$^{4}$Centre for Gravitational Astrophysics, Australian National
  University, Canberra ACT 2601, Australia}
\address{$^{5}$ARC Centre of Excellence for Gravitational Wave Discovery
  (OzGrav), Hawthorn VIC 3122, Australia}
\begin{abstract}
We describe the application of the lattice covering problem to the
 placement of templates in a search for continuous gravitational waves
 from the low-mass X-Ray binary Scorpius X-1.  Efficient placement of
 templates to cover the parameter space at a given maximum mismatch is
 an application of the sphere covering problem, for which an
 implementation is available in the {\LatticeTiling} software library.
 In the case of Sco X-1, potential correlations, in both the prior
 uncertainty and the mismatch metric, between the orbital period and
 orbital phase, lead to complications in the efficient construction of
 the lattice.  We define a shearing {\coord} transformation which
 simultaneously minimizes both of these sources of correlation, and
 allows us to take advantage of the small prior orbital period
 uncertainty.  The resulting lattices have a factor of about 3 fewer
 templates than the corresponding parameter space grids constructed by
 the prior straightforward method, allowing a more sensitive search at
 the same computing cost and maximum mismatch.
\end{abstract}
\maketitle

\section{Introduction}

\ac{ScoX1} is a compact object in a binary system with a low-mass companion
star. \cite{Fomalont2001ScoX1, Steeghs2002ScoX1}
It is believed to be a rapidly spinning neutron star
and a promising source of
continuous gravitational waves \cite{Watts_2008}.  The signal received by an observatory
such as LIGO\cite{LIGO}, Virgo\cite{Virgo} or KAGRA\cite{KAGRA} depends on the parameters of the system,
and a search for that signal loses sensitivity if the incorrect values
are used for those parameters.  Several of the parameters
are uncertain, and one method to ensure that the signal is not missed
is to perform the search at each point in a template bank covering the
relevant parameter space. These include the projected semimajor axis $a_p = a\sin{i}$
of the neutron star's orbit, the orbital period $\Porb$, and the time
$\Tasc$ at which the neutron star crosses the ascending node as measured
in th solar-system barycenter.

The loss of \ac{SNR} associated with an incorrect
choice of parameters is, in a generic Taylor expansion, a quadratic
function of the parameter offsets.  This allows us to write the
fractional loss in SNR, also known as the mismatch, as a squared distance
using a metric on parameter space.  In general, this metric will vary
over the parameter space (i.e., the associated geometry will have
intrinsic curvature), but we can divide the parameter space into small
enough pieces that the space is approximately flat, and the metric can
be assumed to be constant.  In that case, there exists a
transformation to Euclidean {\coord}s.  The problem of placing
templates so that the mismatch of any point in parameter space from
the nearest template is no more than some maximum mismatch $\mu$ is
then equivalent to the problem of covering the corresponding Euclidean
space with spheres of radius $\sqrt{\mu}$.  The most efficient covering
in $n\le 5$ dimensions is the lattice family $\Ans$, which includes
the hexagonal lattice $A_2^*$ and the body-centered cubic lattice
$A_3^*$.  For example, the density of lattice points for $A_4^*$ is a
factor of $2.8$ lower than the corresponding hypercubic
($\mathbb{Z}^4$) lattice.

We use the {\LatticeTiling} module in the LIGO Algorithms Library\cite{lalsuite}
(\texttt{lalsuite}) to investigate efficient lattice coverings for the
parameter space of a search for \ac{ScoX1} using advanced LIGO data.  We
show how the search can be made more efficient by: replacing a
hypercubic grid with an $\Ans$ lattice; accounting for the
elliptical boundaries associated with the correlated prior
uncertainties between orbital period and orbital phase;
defining a sheared {\coord} change such that a particular
combination of the orbital period and orbital phase is unresolved,
and explicitly searching only in the other three dimensions of the
parameter space.  These improvements allow the search to be carried
out using fewer computational resources.  Alternatively, since the
search method we use is tunable, with a trade-off between
computational cost and sensitivity, the more efficient lattice allows
a more sensitive search to be done at the same computing cost.

The plan of this paper is as follows: In \secref{s:CrossCorr}, we
briefly summarize the cross-correlation search for continuous \acp{GW}
as applied to \ac{ScoX1}.  In \secref{s:byhand} we describe the
existing method of template placement ``by hand'' in a rectangular
grid.  In \secref{s:covering} we describe the application of the
sphere covering problem to generation of template lattices, which is
implemented in the {\LatticeTiling} module in the \texttt{lalsuite}
software library\cite{lalsuite}, as described in
\cite{Wette2014_Lattice}.  In \secref{s:paramspace} we consider the
specific features of the parameter space for the \ac{ScoX1} search
which impact our search: \secref{ss:priors} describes the orbital
priors, especially the relationship between orbital period and phase.
In \secref{ss:stdcoords} we consider the standard {\coord}s, where the
time of ascension describing the orbital phase has been propagated
in time to the epoch of the gravitational wave search, inducing prior
correlations between the period and time of ascension.
In \secref{ss:sheared} we show how a shearing transformation can be
used to define a modified period parameter whose prior uncertainty is
independent of the uncertainty on the propagated time of ascension.
In \secref{s:results} we construct a number of lattices and compare
the numbers of templates and modelled computing costs. Finally
\secref{s:conclusions} contains conclusions and implications of this
work.

\section{Background}

\subsection{Cross-Correlation Search for Scorpius X-1}
\label{s:CrossCorr}

The model-based cross-correlation method
\cite{Dhurandhar2007_CrossCorr} has been developed to search for
continuous \acp{GW}, most notably from the low-mass X-ray binary
\ac{ScoX1} \cite{Whelan2015_ScoX1CrossCorr} and applied to mock data
\cite{Messenger2015_MDC1} as well as observational data from Advanced
LIGO's first and second science runs
\cite{LVC2017_O1ScoX1CrossCorr,Zhang2021_O2ScoX1CrossCorr}.  It is a
semi-coherent method where the data are divided into short segments of
duration $200\un{s}\lesssim\Tsft\lesssim 2000\un{s}$, which we call
``SFTs'' because we construct a Short Fourier Transform from each of
them. A detection statistic is constructed including correlations
between pairs of segments separated by a coherence time $\Tmax$ or
less.\footnote{In this paper we describe the original ``demod''
  implementation of the search.  At low frequencies, the search can be
  made more efficient by using resampling to reimplement the loop over
  data and the search over frequencies, as described in
  \cite{Meadors2018_CCResamp}, but the considerations for the template
  bank in the orbital parameters are similar.}  The sensitivity of the
search scales with the number of included pairs; when $\Tmax$ is much
less than the total observation time, the detectable \ac{GW} strain is
proportional to $\Tmax^{-1/4}$.  Since the search is computationally
limited and the computing cost increases with $\Tmax$, the search can
be tuned to trade computational cost for sensitivity.  This tuning can
also be done across the parameter space, with different parameter
space regions being assigned different $\Tmax$ values.  Typically, one
uses more computing resources in regions of parameter space which are
more likely to contain the signal, where the search is inherently more
sensitive, and where it is inherently computationally cheaper.

The output of the search is a detection statistic, which is normalized
to have unit variance.  The value of $\rho$ can then be seen as a
\ac{SNR} for the search.  In the presence of a signal of intrinsic
amplitude $h_0$, the expectation value $\ev{\rho}\propto h_0^2$, with the proportionality
constant being a measure of the sensitivity of the search.
Because the model-based statistic is constructed using signal
parameters such as intrinsic frequency and parameters influencing the
Doppler modulation of the signal, such as sky position and the binary
orbit of the neutron star, the \ac{SNR} in the presence of a
signal will be reduced if the template model parameters differ from
those of the signal. For parameters which are unknown or insufficiently
constrained, the search is run repeatedly at different points in
parameter space to try to find a point close to the true signal. If the
parameter values for a search point are $\{\lambda_i\}$ and the
corresponding true values of the signal are $\{\lambda^s_i\}$, we can
define the \textit{mismatch} $\mu$ as the fractional loss in \ac{SNR}:
\begin{equation}
  \mu = 1 - \frac{\ev{\rho}_{\{\lambda_i\}}}{\ev{\rho}_{\{\lambda^s_i\}}}.
\end{equation}
A Taylor expansion in the $n$ parameters $\{\lambda_i\}$
gives\footnote{This assumes that
  the \ac{SNR} is a local maximum at the true signal point
  $\lambda_i=\lambda^s_i$.  This is not quite true, as shown in
  \cite{Whelan2015_ScoX1CrossCorr}, but it is a good starting point.}
\begin{equation}
  \mu \approx
  \sum_{i=1}^n \sum_{j=1}^n
  g_{ij}(\lambda_i-\lambda_i^s)(\lambda_j-\lambda_j^s),
\end{equation}
where the matrix $\{g_{ij}\}$ acts as a metric on parameter space.
The general form of the metric for the cross-correlation search is
\cite{Whelan2015_ScoX1CrossCorr}
\begin{equation}
  \label{e:crosscorrmetric}
  g_{ij} \approx
  \frac{1}{2}\langle \Delta\Phi_{\alpha,i} \Delta\Phi_{\alpha,j}\rangle_{\alpha}
  \ ,
\end{equation}
where $\alpha$ represents a pair of SFTs, $\Delta\Phi_{\alpha}$ is the
difference in modelled signal phase between the SFTs in the pair,
$\langle\cdot\rangle_{\alpha}$ is an average over SFT pairs weighted
by the antenna patterns and sensitivity of the detectors involved, and
${}_{,i}=\frac{\partial}{\partial\lambda}_i$ is a partial derivative
with respect to the parameter $\lambda_i$.

\begin{figure}[t]
  \begin{center}
     \includegraphics[height=0.42\columnwidth]{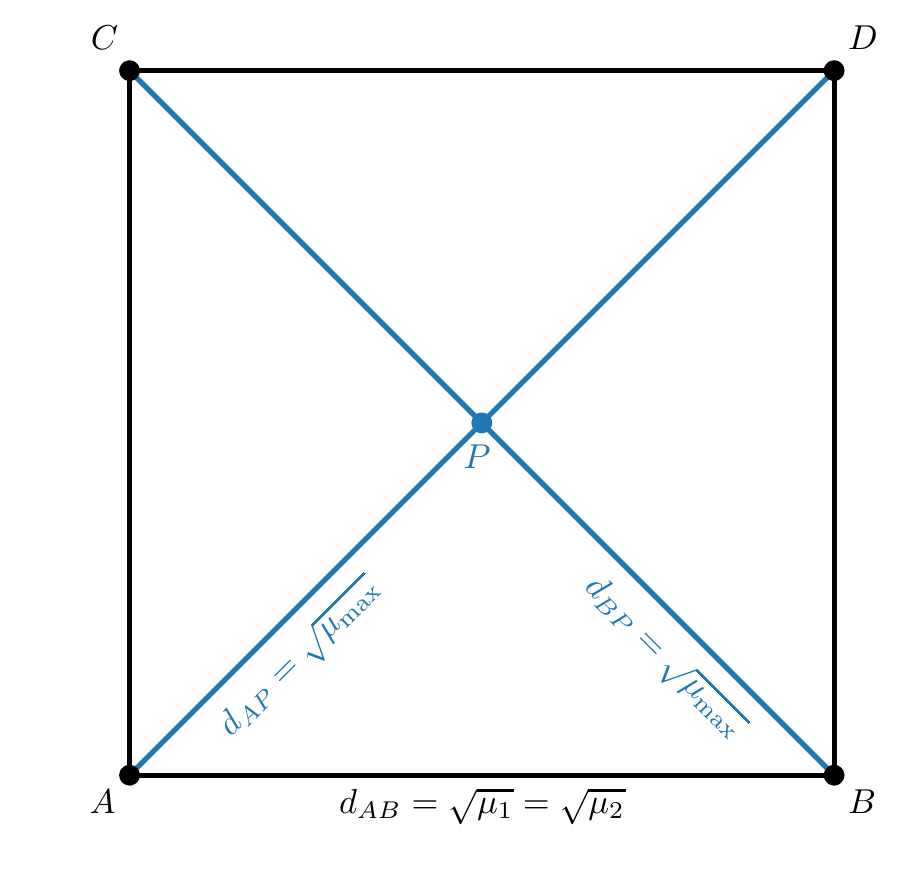}
     \includegraphics[height=0.42\columnwidth]{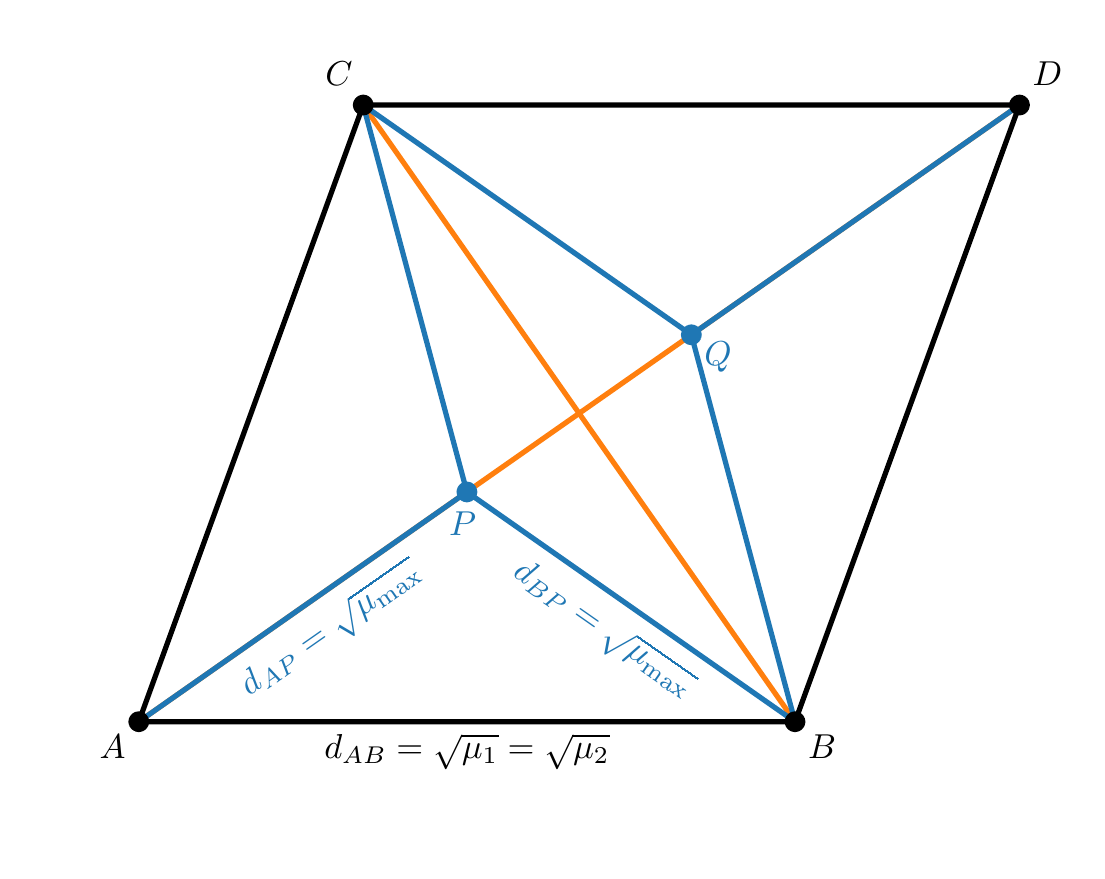}
  \end{center}
  \caption{Illustration of maximum mismatch of the by-hand grid of
    \secref{s:byhand} when the metric is not diagonal.  Specializing
    to the case of $n=2$ and $\mu_1=\mu_2$, we transform to Euclidean
    {\coord}s as described in \secref{s:covering}.  Left: If the
    metric is diagonal, this is just a scaling which transforms the
    rectangle defined by four grid points $ABDC$ into a square with
    sides of length $\sqrt{\mu_1}=\sqrt{\mu_2}$.  The point in
    parameter space farthest from any grid point is the center $P$ of
    the square, with $d_{AP}=d_{BP}=d_{CP}=d_{DP}=\sqrt{\mu_1/2}$.
    Right: If the metric is not diagonal, this square becomes a
    rhombus $ABDC$.  Defining $AD$ to be the long diagonal, the point
    farthest from any vertex is not the center, but a point $P$ on the
    long diagonal with $d_{AP}=d_{BP}=d_{CP}$.  (There is an
    equivalent point $Q$ on the other side of the center with
    $d_{DQ}=d_{BQ}=d_{CQ}$.)  We see that $APB$ (or equivalently $APC$
    or $DQB$ or $DQC$) is an isoceles triangle with
    $d_{AP}=d_{BP}=\sqrt{\mumax}$ and $d_{AB}=\sqrt{\mu_1}$.  In the
    case of a non-diagonal metric, $\angle{APB}$ is an obtuse angle,
    and $d_{AP}=d_{BP}<d_{AB}/\sqrt{2}$, so
    $\mumax>\mu_1/2=(\mu_1+\mu_2)/4$.}
  \label{f:rhombicgrid}
\end{figure}

\subsection{Simple Rectangular Template Placement}
\label{s:byhand}

The cross-correlation analyses run to date examined a parameter space
divided up into rectangular regions, small enough to
assume a constant metric. Then, a set of discrete points is placed
over the parameter space which lie on a rectangular grid with
spacing $\delta\lambda_i$ in the $\lambda_i$ direction, using what we
refer to as the ``by hand'' method.  The number of points used is
\begin{equation}
  N_i =
  \left \lceil
    \frac{\lambda_i^{\text{max}} - \lambda_i^{\text{max}}}{\delta \lambda_i},
  \right \rceil
\end{equation}
where $\lceil\cdot\rceil$ indicates rounding to up to the next
integer.  The spacing $\delta\lambda_i$ is chosen to be
\begin{equation}
  \delta \lambda_i = \sqrt{\frac{\mu_i}{g_{ii}}}
\end{equation}
so that the mismatch between adjacent points\footnote{Note that for
  historical reasons, $\mu_i$ is defined as the mismatch between
  adjacent points in the grid, rather than the maximum mismatch
  between some point in the parameter space and the nearest grid
  point.  This is the origin of the factor of $\frac{1}{4}$ appearing
  in \eqref{e:mumaxgrid}.} in the $\lambda_i$ direction is
$g_{ii}(\delta \lambda_i)^2=\mu_i$.  If the metric is approximately
diagonal, $g_{ij}=g_{ii}\delta_{ij}$, then the point in the parameter
space farthest (in the sense of the metric) from any grid point is
$\frac{\delta \lambda_i}{2}$ away in the $\lambda_i$ direction, and
has a total mismatch of
\begin{equation}
  \label{e:mumaxgrid}
  \mumax = \sum_{i=1}^n g_{ii} \left(\frac{\delta \lambda_i}{2}\right)^2
  = \frac{1}{4} \sum_{i=1}^n \mu_i
\end{equation}
If the metric is not diagonal, the procedure described above will lead
to a maximum mismatch greater than that given in \eqref{e:mumaxgrid},
as illustrated in \figref{f:rhombicgrid}.  This approach is
conservative and can result in much larger template banks if the
metric contains large correlations. The number of templates could be
reduced by accounting for the metric correlations, which will be
discussed later in \secref{ss:sheared}.

\subsection{Covering Lattices}
\label{s:covering}

The general problem of choosing a set of template points with a prescribed
maximum mismatch distance $\mumax$ between any point in the parameter
space and the nearest template is an application of the
\textit{sphere covering problem} \cite{conway1998sphere}.  Since we
treat the metric $\{g_{ij}\}$ as approximately constant, there is
always a linear transformation of the parameters $\{\lambda_i\}$ into
Euclidean coordinates $\{x_i\}$; the mismatch between two
points separated by parameter differences $\{\Delta\lambda_i\}$ is then
\begin{equation}
  \sum_{i=1}^n \sum_{j=1}^n g_{ij} (\Delta\lambda_i)(\Delta\lambda_j)
  = \sum_{i=1}^n (\Delta x_i)^2
  \ .
\end{equation}
The template placement problem is then simplified to  one of placing
(hyper-)spheres of radius
$\sqrt{\mumax}$ in the $\{x_i\}$ space so that every point of the
region of interest is covered by at least one sphere.  To efficiently
cover the space, the overlap between spheres should be minimized. This
is quantified using the normalized thickness or center density $\theta$,
which is the average number of templates per unit volume for the
unit sphere.

A sphere covering based on a repeating pattern is known as a lattice.
The number of templates required to cover the space will be at a minimum
when the lattice has the smallest thickness $\theta$. A perfect lattice
has a thickness of 1.

The simplest lattice is the cubic lattice $\Zn$, which has
points equally spaced in each of the (Euclidean) coordinate
directions.  The ``by hand'' lattice of \secref{s:byhand} is an
example of a $\Zn$ lattice, if the metric $\{g_{ij}\}$ is
diagonal and all of the mismatches $\mu_i$ are chosen to be equal.  A
more efficient lattice is $\Ans$, which is a general analogue of the
hexagonal lattice.  For the sphere
covering problem, the thinnest lattice is the $\Ans$ lattice, which
in two dimensions has a hexagonal principal cell. The principal cell is the
set of points closest to given point in a lattice, and the vertices
are locations where covering spheres intersect.
It has been shown for $n\le 5$ that $\Ans$ is the
most efficient covering lattice, i.e., has the smallest thickness
$\theta$ \cite{conway1998sphere}, and for higher
dimensions it is typically close to the most efficient
covering \cite{Prix2007_Lattice}. Since a more efficient lattice
allows the same volume of parameter space to be covered with fewer
templates, it can reduce the necessary computing cost at a given
sensitivity.\footnote{But see \cite{Allen2021Lattice}.}

Construction of $\Zn$ and $\Ans$ lattices in physical coordinates
$\{\lambda_i\}$ given a constant mismatch metric $\{g_{ij}\}$ is
implemented by the {\LatticeTiling} module in the
\texttt{lalsuite} software library\cite{lalsuite}, as described
in \cite{Wette2014_Lattice}.  A particular challenge is ensuring that
the area within the boundaries of a search region is completely covered,
which sometimes requires retaining templates whose parameters lie
outside the search region.  We shall see that this can necessitate
some care in choosing coordinates to take advantage of underresolved
directions in parameter space.

\section{Parameter Space for Sco X-1 Search}
\label{s:paramspace}

\subsection{Observational Priors}

\label{ss:priors}

The \ac{GW} signal produced by a spinning neutron star is
the system is nearly periodic in the neutron star's rest frame, and
Doppler shifted as a result of the motion of the detector as the Earth
rotates and moves in its orbit and, in the case of a low-mass X-ray
binary such as \ac{ScoX1}, of the neutron star in its own orbit with
its binary companion.  For an accreting neutron star in approximate
spin equilibrium, the frequency $f_0$ can be approximated as
constant.\footnote{In practice, this equilibrium will be imperfect,
  leading to some ``spin wandering'', but the impact of deviations
  from equilibrium was shown in \cite{Whelan2015_ScoX1CrossCorr} to be
  limited when the coherence time is not too long, especially with the
  levels of spin wandering predicted by
  \cite{Mukherjee2018_Spinwander}.}  The Doppler shift from detector
motion is primarily affected by sky position, which for \ac{ScoX1} is
well enough known\cite{2mass06} that its uncertainty does not affect
the search.  The Doppler shift from the binary motion is affected by
five orbital parameters: eccentricity, orientation, projected orbital
speed, orbital period, and orbital phase\cite{Leaci2015_ScoX1}. The
orbit of \ac{ScoX1} is believed to be nearly
circular\cite{Galloway2014}, so that the search needs to cover only
three orbital parameters: projected speed, period, and
phase.

\begin{figure}[t]
  \begin{center}
    \includegraphics[width=0.47\columnwidth]{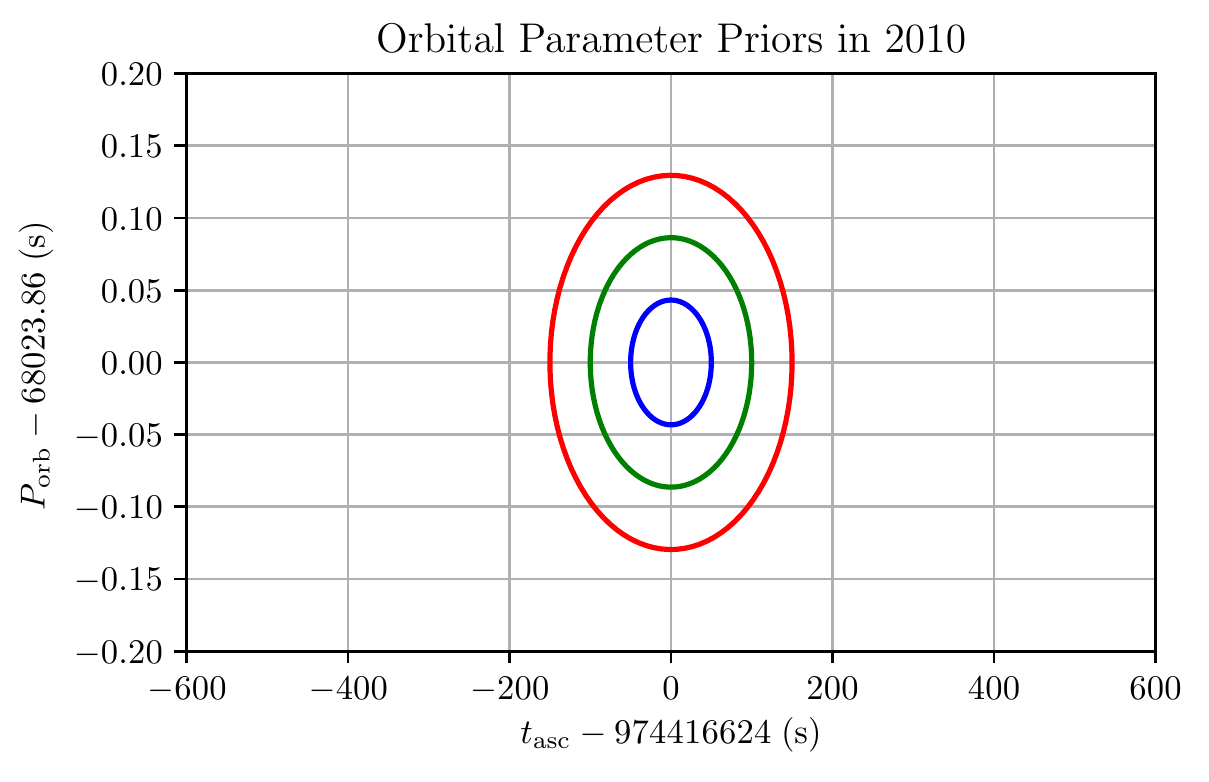}
    \includegraphics[width=0.47\columnwidth]{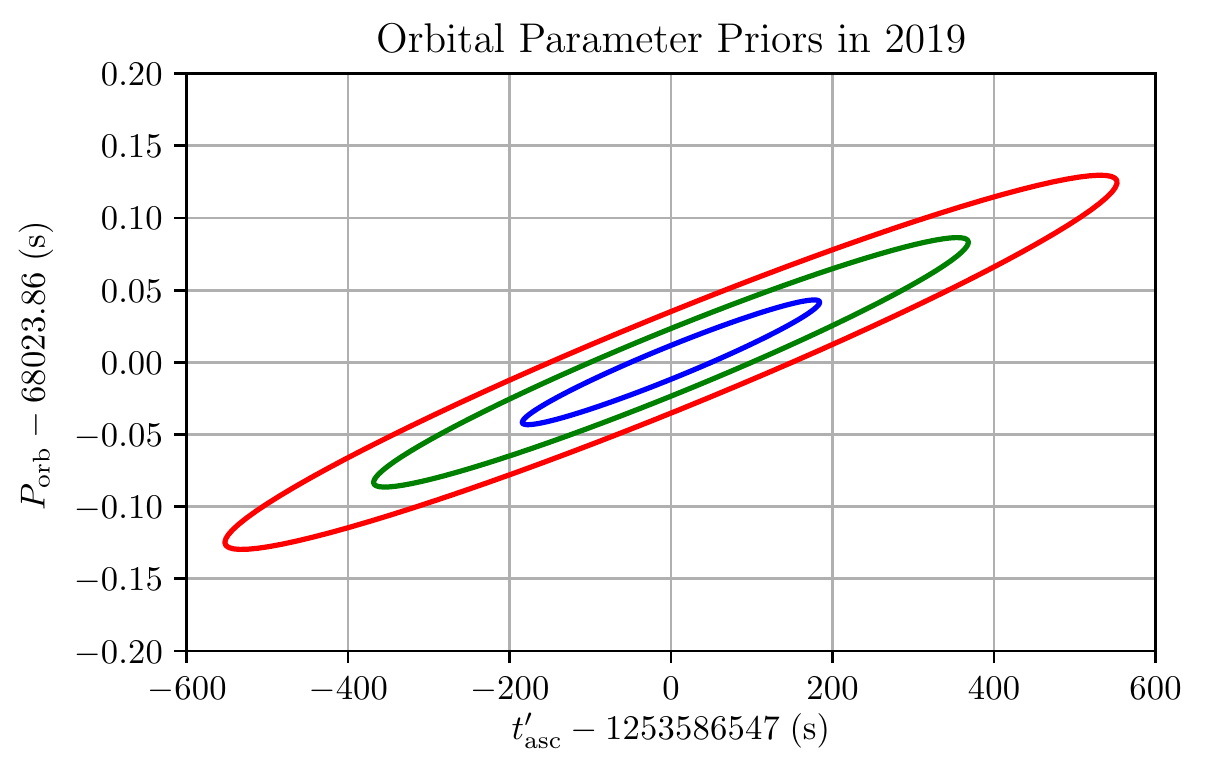}
  \end{center}
  \caption{Orbital parameter constraints from \cite{Wang2018_PEGS3}.
    If the time of ascension $\TascThen$ is quoted in 2010 (left
    panel) its uncertainty is uncorrelated with the orbital period
    $\Porb$.  If we propagate forward by $\norb$ orbits to determine
    an equivalent time of ascension
    $\TascNow = \TascThen + \norb\Porb$, this introduces correlations.
    In each case the level surfaces of the probability distribution
    are shown, for which $\chi^2$, defined in \eqref{e:chisqThen} and
    \eqref{e:chisqNow} equals $1^1$, $2^2$ and $3^2$.  We refer to
    these as at normalized distances of $1\sigma$, $2\sigma$, and
    $3\sigma$, and they correspond to cumulative probabilities of
    {\pctonesig\%}, {\pcttwosig\%}, and {\pcttresig\%},
    respectively.}
  \label{f:TPprior}
\end{figure}

The best constraints on these come from \cite{Wang2018_PEGS3}.  The constraint on orbital period $\Porb$ is Gaussian, with a mean
of $\BestPorb=\WangPorbSec\un{s}$ and a standard deviation of
$\sigPorb=\WangdPorbSec\un{s}$.  The orbital phase is described by
time of ascension $\Tasc$, which is the time at which the neutron star
crosses the plane of the sky moving away from the observer (i.e.,
crosses the ascending node).  The constraint on this is also Gaussian,
with a mean of
$\BestTascThen=\WangTascGPS~ \text{GPS}~$({\WangTascDayTime})
and a standard deviation of $\sigTascThen=\WangdTascGPS\un{s}$.  These
estimates are uncorrelated, as shown in the left panel of \figref{f:TPprior},
but if we
convert the time of ascension to a subsequent equivalent time
$\TascNow=\TascThen+\norb\Porb$, a correlation is induced, as
described in \secref{ss:stdcoords} and shown in the right panel of
\figref{f:TPprior}.  The constraints on the orbital
velocity of the neutron star in \cite{Wang2018_PEGS3} are described in
terms of the amplitude of the component of velocity along the line of
sight, known as $K_1$, and consist of constraints that
$40\un{km/s}\le K_1\le 90\un{km/s}$, but without a well-determined
probability density between those limits.  Searches for \acp{GW} from
\ac{ScoX1} typically use a uniform prior distribution on this
parameter.  The parameter used is also typically written as
the line-of-sight component of the semimajor axis of the orbit,
$\ap=\frac{K_1\Porb}{2\pi}$.  Since the relative uncertainty on
$\Porb$ is much less than on $K_1$, one assumes a uniform prior on
$\ap$ for $\WangAsiniMin\un{lt-s}\le\ap\le\WangAsiniMax\un{lt-s}$, where
the units on $\ap$ are given in light-seconds.

\subsection{Standard Search {\Coord}s}
\label{ss:stdcoords}

The phase derivatives $\{\Delta\Phi_{\alpha,i}\}$ appearing
in \eqref{e:crosscorrmetric} are computed in
\cite{Whelan2015_ScoX1CrossCorr} for the standard search coordinates
$\{\lambda_i\}\equiv\{f_0,\ap,\TascNow,\Porb\}$.  Here we introduce the
$\TascNow$ as the time of ascension at a given point in the
propagated 2019 coordinates, indicated by the prime. For long searches
which evenly sample the orbital phase of the binary, the
non-negligible metric elements have the approximate form\footnote{Note
  that the original formula for $\gNowTP$--(4.20h) in
  \cite{Whelan2015_ScoX1CrossCorr}--contains a sign error which has
  not been relevant previously because approximate form of $\gNowTP$
  has only been used to set it to zero.}
\begin{subequations}
  \begin{gather}
    {\gff}
    \approx 2\pi^2
    \left\langle
      \Tdiff_\alpha^2
    \right\rangle_{\!\alpha}
    \approx \frac{2\pi^2}{3}\Tmax^2,
    \\
    {\gaa} = 4\pi^2 f_0^2
    \left\langle
      \sin^2\frac{\pi \Tdiff_\alpha}{\Porb}
    \right\rangle_{\!\alpha}
    \approx
    2\pi^2 f_0^2
    \left(
      1 - \sinc \frac{2\Tmax}{\Porb}
    \right),
    \\
    \begin{pmatrix}
      {\gNowTT} & {\gNowTP} \\ {\gNowPT} & {\gNowPP}
    \end{pmatrix}
    \approx
    \begin{pmatrix}
      1
      &
      \frac{
        -(\TascNow -
        \left\langle
          \Tbar_{\alpha}
        \right\rangle_{\!\alpha})
      }
      {\Porb}
      \\
      \frac{
        -(\TascNow -
        \left\langle
          \Tbar_{\alpha}
        \right\rangle_{\!\alpha})
      }
      {\Porb}
      &
      \frac{
        \left\langle
          (\TascNow-\Tbar_{\alpha})^2
        \right\rangle_{\!\alpha}
      }
      {\Porb}
    \end{pmatrix}
    \frac{16\pi^4 f_0^2 a_p^2}{\Porb^2}
    \left\langle
      \sin^2\frac{\pi \Tdiff_\alpha}{\Porb}
    \right\rangle_{\!\alpha},
  \end{gather}
\end{subequations}
where $\Tdiff_\alpha$ is the difference between the timestamps of the
two SFTs in pair $\alpha$, and $\Tbar_\alpha$ is their mean.  Note
that the implementation in \texttt{lalsuite} \cite{lalsuite} uses the
exact metric elements, which include additional (generally small)
off-diagonal elements.

If we define the midpoint of the run (according to the weighted
average $\langle\cdot\rangle_{\alpha}$) to be
$\muobs = \langle\Tbar_{\alpha}\rangle_{\alpha}$ and the variance as
$\sigobs^2 = \langle(\Tbar_{\alpha}-\muobs)^2\rangle_{\alpha}$ The
metric elements on the $\TascNow,\Porb$ subspace become
\begin{subequations}
  \begin{gather}
    {\gNowTT} \approx \frac{16\pi^4 f_0^2 a_p^2}{\Porb^2}
    \left\langle
      \sin^2\frac{\pi \Tdiff_\alpha}{\Porb}
    \right\rangle_{\!\alpha},
    \\
    {\gNowTP} \approx
    \left(
      \frac{-(\TascNow-\muobs)}{\Porb}
    \right)
    {\gNowTT},
    \\
    {\gNowPP} \approx
    \left(
      \frac{(\TascNow-\muobs)^2+\sigobs^2}{\Porb^2}
    \right)
    {\gNowTT}.
  \end{gather}
\end{subequations}
Note that, as shown in \cite{Whelan2015_ScoX1CrossCorr}, if we ignore
any data gaps and the noise and antenna pattern weighting, for
an observing run of duration $\Trun$ and coherence time $\Tmax$,
$\sigobs^2\approx\frac{\Trun^2}{12}$, and
$\left\langle \sin^2\frac{\pi \Tdiff_\alpha}{\Porb}
\right\rangle_{\!\alpha}\approx\frac{1}{2}\left( 1 - \sinc
  \frac{2\Tmax}{\Porb} \right)$.

If we choose $\norb$ so that
\begin{equation}
  \BestTascNow = \BestTascThen + \norb\BestPorb
\end{equation}
is as close as possible to $\muobs$, we can minimize the magnitude of
\begin{equation}
  {\gNowTP} \approx
  \left(
    \frac{\BestTascThen + \norb\BestPorb-\muobs}{\BestPorb}
  \right)
  {\gNowTT}.
\end{equation}
This achieved by taking
\begin{equation}
  \label{e:norb-unsheared}
  \norb = \nint{\frac{\muobs-\BestTascThen}{\BestPorb}}
\end{equation}
where $\nint{\cdot}$ indicates rounding to the nearest integer.

To give a concrete example, we consider the LIGO-Virgo O3 data run
\cite{LVK2020_ObsScenarios} which began on {\OiiiaStartDayTime} (GPS
{\OiiiaStartGPS}), continued until a commissioning break at
{\OiiiaEndDayTime} (GPS {\OiiiaEndGPS}), resumed on
{\OiiibStartDayTime} (GPS {\OiiibStartGPS}), and ended on
{\OiiibEndDayTime} (GPS {\OiiibEndGPS}).
Neglecting variability of antenna patterns and noise spectra, as well
as any data gaps other than the commissioning break, we find an
average time of $\muobs=\text{GPS}~\muObsGPS\equiv${\muObsDayTime}.
This translates into an optimal $\norb=\bestNorb$, corresponding to
$\BestTascNow=\text{GPS}~\bestTascGPS\equiv${\bestTascDayTime}.\footnote{The
  actual values for O3 including duty cycle, noise weighting and
  antenna patterns, and using the exact form of the metric, will be
  slightly different, but we will use the values above for
  illustration in this paper.}

\begin{figure}[t]
  \begin{center}
    \includegraphics[width=0.47\columnwidth]{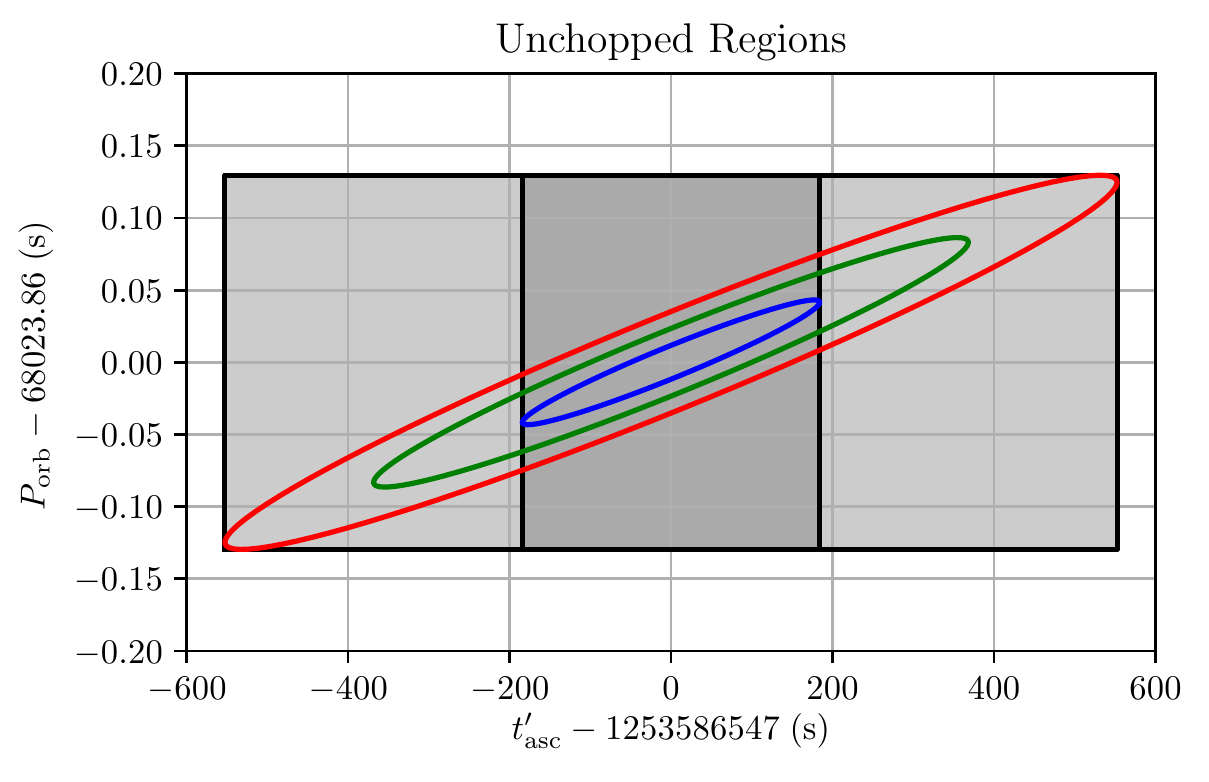}
    \includegraphics[width=0.47\columnwidth]{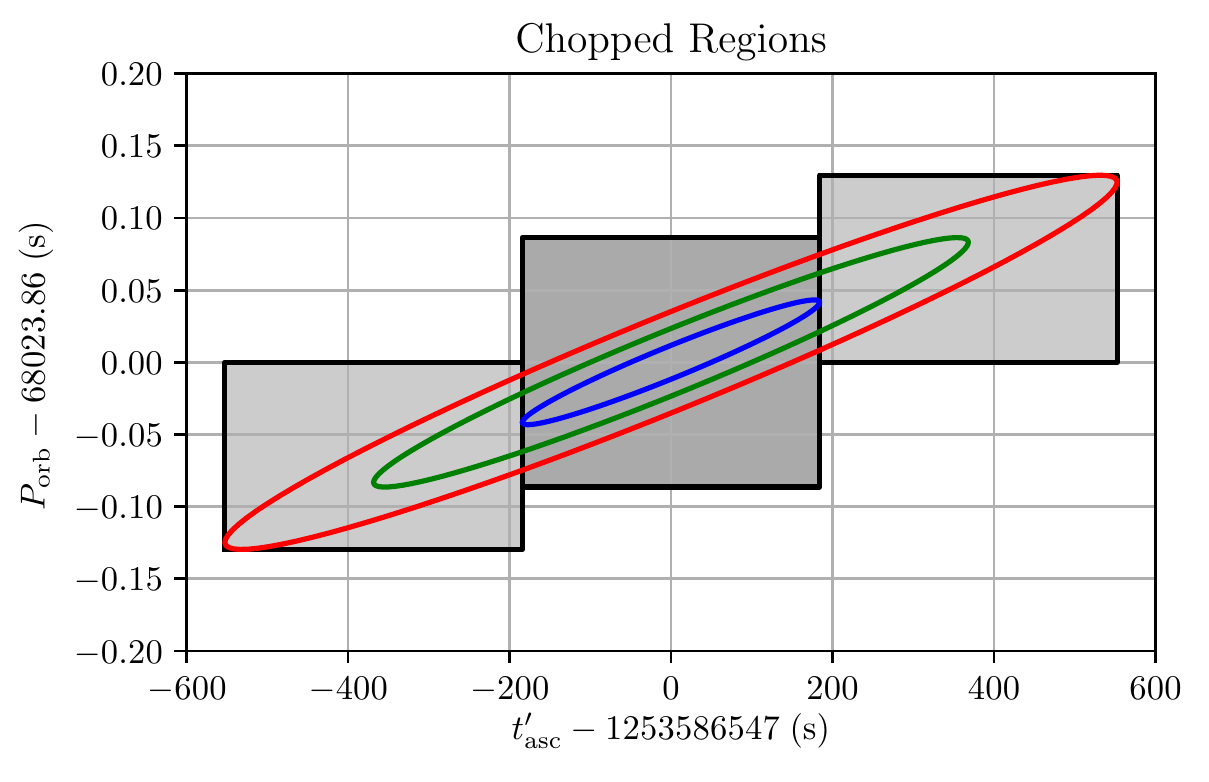}
  \end{center}
  \caption{Left: rectangular search region boundaries cover the entire $\TascNow,\Porb$
    region, as used in O1 CrossCorr \cite{LVC2017_O1ScoX1CrossCorr}. The darker
    region bounded by one-sigma in the $\Tasc$ direction indicates a higher
    likelihood of finding the signal in that region of parameter space. Right:
    Rectangular regions more closely concentrated on the uncertainty ellipses
    than the O1 search regions. The areas in $\TascNow,\Porb$ have been ``chopped''
    to eliminate extra parameter space area where a signal is not likely to be found. }
  \label{f:TP_regions}
\end{figure}

The joint prior on $\TascNow$ and $\Porb$ will remain a multivariate
Gaussian, but now with a non-diagonal variance-covariance matrix.  The
marginal prior on $\TascNow$ will be a Gaussian with mean
$\BestTascNow$ and variance
\begin{equation}
  \sigTascNow^2 = \sigTascThen^2 + \norb^2\sigPorb^2.
\end{equation}
The joint prior can be illustrated by plotting level curves of the
quantity
\begin{equation}
  \label{e:chisqThen}
  \chi^2
  =
  \left(
    \frac{\TascThen-\BestTascThen}{\sigTascThen}
  \right)^2
  +
  \left(
    \frac{\Porb-\BestPorb}{\sigPorb}
  \right)^2,
\end{equation}
whose prior distribution is a chi-squared with two degrees of freedom
(\figref{f:TPprior}, right panel).  A bit of algebra shows that
\begin{equation}
  \label{e:chisqNow}
  \chi^2
  =
  \frac{\sigTascNow^2}{\sigTascThen^2}
  \left[
    \left(
      \frac{\Porb - \BestPorb}{\sigPorb}
    \right)^2
    - \frac{2\norb\sigPorb}{\sigTascNow}
    \left(
      \frac{\TascNow - \BestTascNow}{\sigTascNow}
    \right)
    \left(
      \frac{\Porb - \BestPorb}{\sigPorb}
    \right)
    +
    \left(
      \frac{\TascNow - \BestTascNow}{\sigTascNow}
    \right)^2
  \right].
\end{equation}
In previous searches, rectangular boundaries have been used in all
coordinate directions in the parameter space. For the O1 search, these
regions covered out to $3\sigma$ of the marginal priors on $\TascNow$
and $\Porb$, as shown in Figure~1 of \cite{LVC2017_O1ScoX1CrossCorr}.
If we use a similar approach in O3 (\figref{f:TP_regions}, left panel),
the search regions cover a large area of $\TascNow,\Porb$
parameter space with negligible prior probability.  Since the middle
third of the $\TascNow$ range is searched separately (at a higher
coherence time $\Tmax$, since the prior probability density is higher
there), a simple approach can reduce the over-coverage of the search region.
The $\Porb$ search range is different for each of the rectangular
regions covering different ranges of $\TascNow$, discarding regions in
which the prior $\chi^2\gtrsim 3^2$.  These ``chopped'' regions are
shown in the right panel of \figref{f:TP_regions}.
The chopped regions can be achieved without significant
modification to the previously existing search code.

\begin{figure}[t]
  \begin{center}
    \includegraphics[width=0.47\columnwidth]{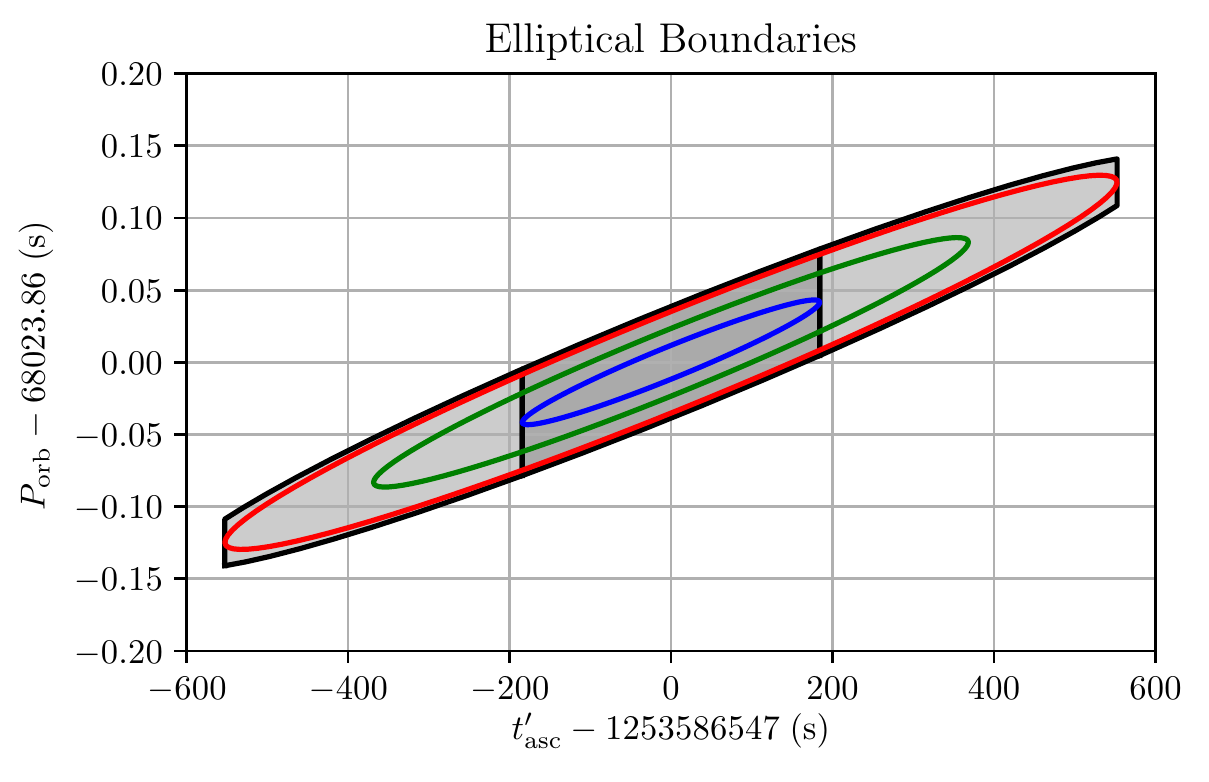}
    \includegraphics[width=0.47\columnwidth]{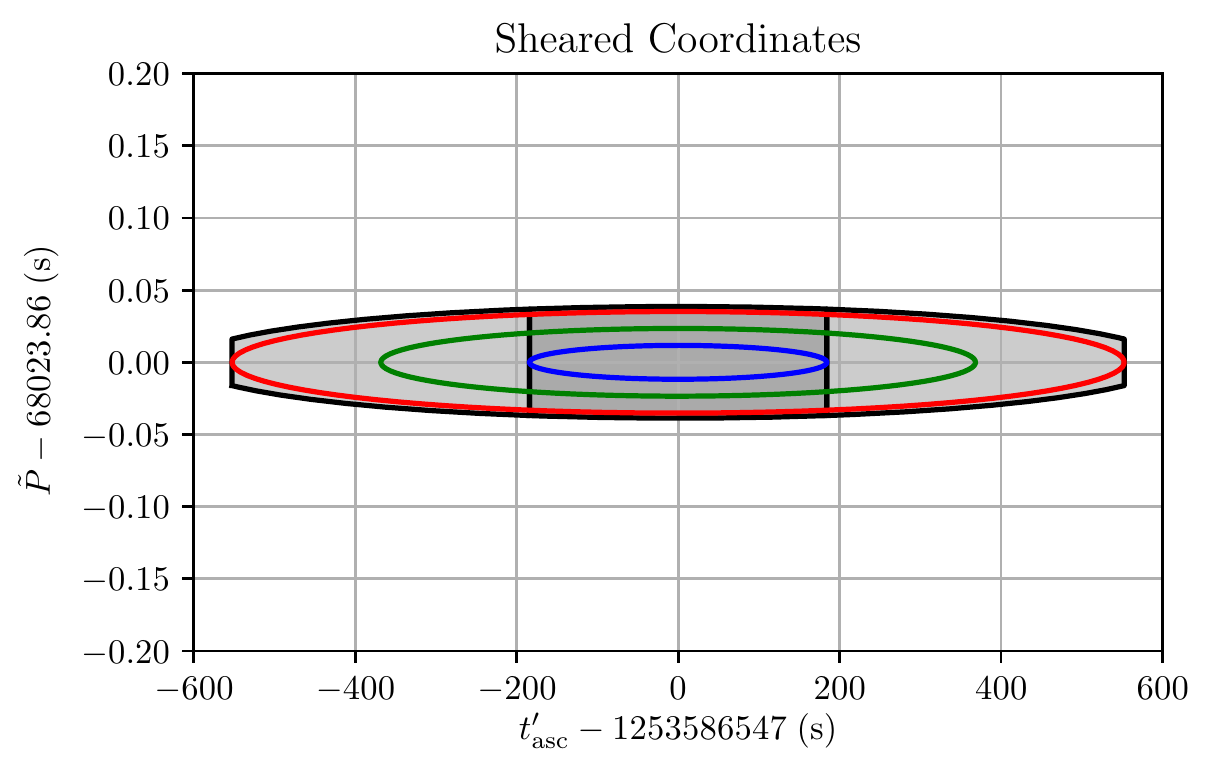}
  \end{center}
  \caption{Left: Search region boundaries in $\TascNow,\Porb$ now
    define a boundary function, where the range of $\Porb$ is computed
    as a function of $\TascNow$, defined in
    \eqref{e:PorbEllipticalBoundary}.  We choose $k=3.3$ for the
    boundary.  For reference, the colored ellipses are level surfaces
    at $1\sigma$, $2\sigma$, and $3\sigma$, i.e., $\chi^2=1^2$, $2^2$,
    and $3^3$, as defined in \eqref{e:chisqNow}.  The darker shaded
    region lies within $\pm 1\sigma$ of the marginal distribution on
    $\TascNow$, where the signal is more likely to be found, and the
    lighter-shaded region is between $\pm 3\sigma$ and $\pm 1\sigma$.
    Dividing up the search regions based on $\TascNow$ rather than
    $\chi^2$ is more efficient since $\TascNow$ is always resolved,
    and $\Porb$ may not be.  Right: the same search regions in a
    ``sheared'' set of {\coord}s $\TascNow,\PorbShear$, where
    $\PorbShear$ is a linear combination of $\Porb$ and $\TascNow$,
    defined in \eqref{e:PorbShear}, which aligns the constant-$\chi^2$
    ellipses with the {\coord} axes.}
  \label{f:TP_regions_ellip}
\end{figure}

To further
improve the efficiency of the parameter space coverage, we can define
an elliptical boundary function which sets the range of $\Porb$
continuously as a function of $\TascNow$.  This function can be used
in the {\LatticeTiling} module to restrict template placement to
those needed to cover the prior ellipse corresponding to
$\chi^2\le k^2$ for a particular $k$:
\begin{equation}
  \label{e:PorbEllipticalBoundary}
    \frac{\Porb - \BestPorb}{\sigPorb}
    \in
    \frac{\norb\sigPorb}{\sigTascNow}
    \left(
      \frac{\TascNow - \BestTascNow}{\sigTascNow}
    \right)
    \pm
    \frac{\sigTasc}{\sigTascNow}
    \sqrt{
      k^2
      -
      \left(
        \frac{\TascNow - \BestTascNow}{\sigTascNow}
      \right)^2
    }.
\end{equation}
This is used to define a search region, together with a constant
boundary on $\TascNow$:
\begin{equation}
  \BestTascNow - k\sigTascNow \le
  \TascNowMin \le \TascNow \le \TascNowMax
  \le \BestTascNow + k\sigTascNow,
\end{equation}
and illustrated in the left panel of \figref{f:TP_regions_ellip}.
Note that we choose $k=3.3$ rather than $k=3$ as the boundary, since
the former encloses {\pcttrepttresig\%} of the prior probability, while
the latter would enclose only {\pcttresig\%}.

\subsection{Sheared {\Coord}s}
\label{ss:sheared}

The joint prior uncertainty in $\TascNow$,$\Porb$ space complicates the
placement of lattice points neatly in coordinate directions. The fact
that the semimajor axis of the uncertainty ellipses does not lie in a
coordinate direction forces rows of lattice points calculated from a
diagonal metric to be placed over a complicated area in parameter
space, which is illustrated in \secref{s:results}. A coordinate
transformation can be performed that
preserves the diagonal metric and shears the coordinates from
$(\TascNow,\Porb)$ to $(\TascNow,\PorbShear)$, aligning the semimajor axis of
the uncertainty ellipses with the coordinate directions,
as shown in the right panel of \figref{f:TP_regions_ellip}.
The lattice points are then chosen in a straightforward way, before a
transformation is then performed back to the physical coordinates.  In
particular, this simplifies the question of whether multiple templates
are necessary to cover the period direction.  Looking at the right
panel of \figref{f:TPprior} or the left panel of
\figref{f:TP_regions_ellip}, we see that the marginal uncertainty in
$\Porb$ is considerably larger than the conditional uncertainty at a
particular value of $\TascNow$.  Changing {\coord}s to $\PorbShear$,
which is observationally uncorrelated with $\TascNow$, allows us to
cover a range of period values corresponding to this smaller marginal
uncertainty.

We can accomplish this {\coord} transformation by subtracting from
$\Porb$ the centerline of the observational uncertainty ellipse and
defining
\begin{equation}
  \label{e:PorbShear}
  \PorbShear =
  \Porb
  -
  \frac{\norb\sigPorb}{\sigTascNow}
  \left(
    \frac{\TascNow - \BestTascNow}{\sigTascNow}
  \right)
  \sigPorb,
\end{equation}
so that
\begin{equation}
  \label{e:chisqShear}
  \chi^2
  =
  \left(
    \frac{\TascNow-\BestTascNow}{\sigTascNow}
  \right)^2
  +
  \left(
    \frac{\PorbShear-\BestPorb}{\sigPorbShear}
  \right)^2,
\end{equation}
and the priors on $\TascNow$ and $\PorbShear$ are once again
independent Gaussians.  Note that
\begin{equation}
  \sigPorbShear = \left(\frac{\sigTascThen}{\sigTascNow}\right) \sigPorb,
\end{equation}
so the area of the uncertainty ellipse is the same in all three sets
of {\coord}s: $(\TascThen,\Porb)$, $(\TascNow,\Porb)$, and
$(\TascNow,\PorbShear)$.

This transformation affects the metric:
\begin{subequations}
  \begin{gather}
    \gShearTT = \gNowTT + 2 \dPdTShear\gNowTP + \dPdTShear^2 \gNowPP, \\
    \gShearTP = \gNowTP + \dPdTShear\gNowPP, \\
    \gShearPP = \gNowPP,
  \end{gather}
\end{subequations}
where
\begin{equation}
  \dPdTShear = \norb \left(\frac{\sigPorb}{\sigTascNow}\right)^2
  = \norb \frac{\sigPorb^2}{\sigTascThen^2 + \norb^2\sigPorb^2}.
\end{equation}
In order to make the metric as close to diagonal as possible in these
{\coord}s, we should choose a different $\norb$ from that defined in
\eqref{e:norb-unsheared}.  Instead we make
\begin{equation}
  \gShearTP \approx
  \left(
    \norb - \frac{\muobs-\BestTascThen}{\BestPorb}
    + \norb  \left(\frac{\sigPorb}{\sigTascNow}\right)^2
    \frac{(\BestTascNow-\muobs)^2+\sigobs^2}{\BestPorb^2}
  \right)
  {\gNowTT}
\end{equation}
close to zero.  If we set this to zero and solve algebraically for
$\norb$, we get
\begin{equation}
  \label{e:norb-sheared}
  \norb \approx \frac{\muobs - \BestTascThen}{\BestPorb}
  \left( 1 + \left(\frac{\sigPorb}{\sigTascNow}\right)^2 \
  \frac{(\BestTascNow- \muobs)^2 + \sigobs^2}{\BestPorb^2} \right)^{-1}.
\end{equation}
Since the definitions of $\BestTascNow$ and $\sigTascNow$
depend on $\norb$ as well, we need to solve iteratively for the
optimal $\norb$ to minimize the metric correlation in these sheared
{\coord}s.  This converges quickly, giving, for the reference values
used in this paper, $\norb = \shearBestNorb$, corresponding to
$\BestTascNow=\text{GPS}~\shearBestTascGPS
\equiv${\shearBestTascDayTime}.\footnote{Again, the
  actual best value using the data with gaps, antenna patterns and
  variable noise level, as well as the exact metric, will be slightly
  different, but the relationship between the choices of $\norb$
  optimized for sheared and unsheared {\coord}s is illustrative.}
With this choice, we have {\coord}s $\TascNow$ and $\PorbShear$ with
no prior correlation and negligible correlation in the search metric.

\begin{figure}[t]
  \begin{center}
    \includegraphics[width=\columnwidth]{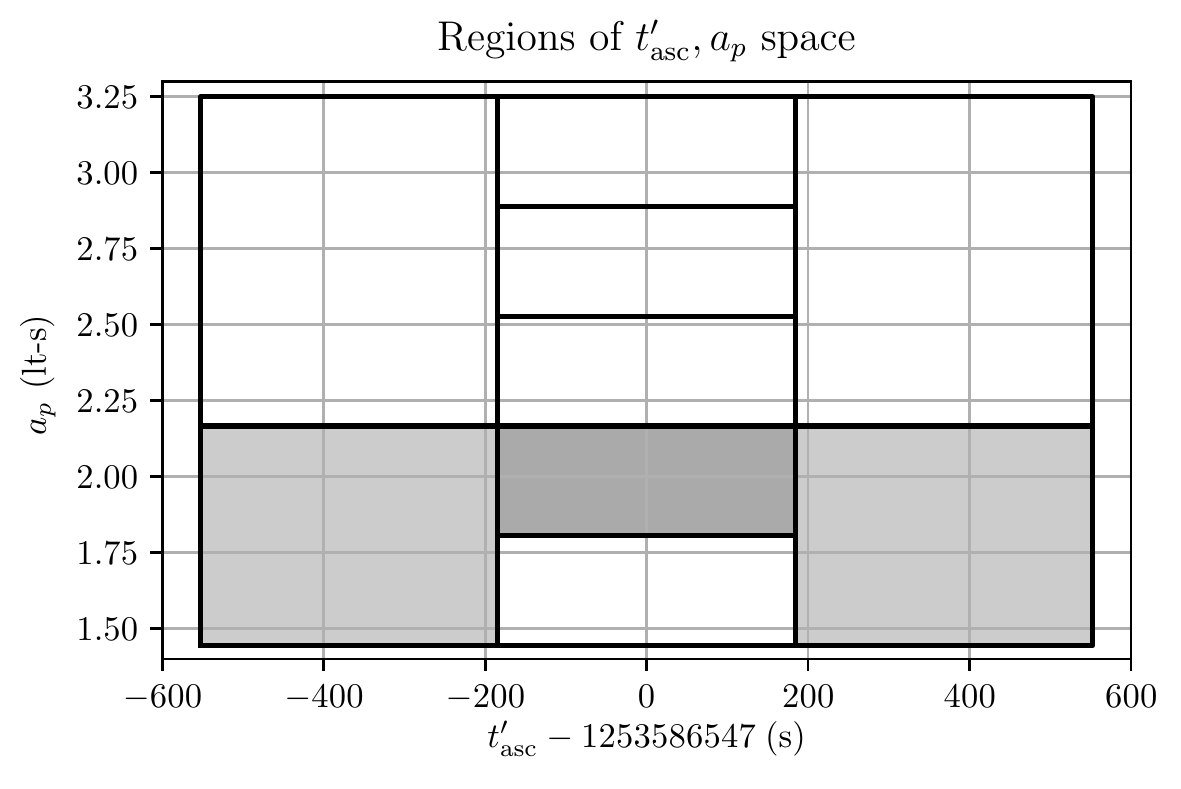}
  \end{center}
  \caption{Cells of $\TascNow,\ap$ space for sample lattice
    construction.  Each rectangular cell has its own coherence time
    $\Tmax$, corresponding to a coherence time used in
    \cite{LVC2017_O1ScoX1CrossCorr}, and we construct a lattice in
    each of these cells.  The range of orbital period values for each
    cell is a function of $\TascNow$, as illustrated in
    \figref{f:TP_regions} or \figref{f:TP_regions_ellip}.  We
    construct the lattice in all 9 of these regions, and include the
    template counts in the computing cost estimate.  For the three
    shaded regions, we also include the templates in the corresponding
    $\TascNow,\Porb$ or $\TascNow,\PorbShear$ plot of the lattice. }
  \label{f:Ta_regions}
\end{figure}

\section{Example Lattices and Results}
\label{s:results}

To quantify the reduction in number of search templates and computing
costs at a given mismatch, we construct sample lattices of each type
for a variety of representative regions in parameter space.  For each
choice of {\coord} system and lattice type, we construct $9\times 14$
lattices, corresponding to the nine regions of orbital parameter space
$(\TascNow,\Porb,\ap)$ or $(\TascNow,\PorbShear,\ap)$ shown in
\figref{f:Ta_regions} and fourteen frequency bands beginning at
$25\un{Hz}$ and ending at $2000\un{Hz}$.  Each of these regions has
its own $\Tmax$ value taken from the search in
\cite{LVC2017_O1ScoX1CrossCorr}.
In that search, the frequency $f_0$ was split into
ranges of width $0.05\un{Hz}$, and a search job covered that range of
frequencies along with one of the orbital parameter space regions.
Rather than constructing the full set of $9\times 39500$ lattices
covering all the bands from $25\un{Hz}$ to $2000\un{Hz}$, we choose
one $0.0005\un{Hz}$ range from the middle of each band, construct the
nine lattices (one for each orbital parameter cell) corresponding to
that range, and scale up the number of templates by the number of such
ranges in the band.  Since the computing cost scales roughly with the
number of templates times the number of SFT pairs, we approximate the
computing cost for each band $i$ and cell $c$ as
$N^{\text{pair}}_{ic} N^{\text{tmplt}}_{ic}$.  We estimate the number
of SFT pairs as in \cite{Whelan2015_ScoX1CrossCorr} by
\begin{equation}
N^{\text{pair}}_{ic} \approx N_{\text{det}}^2\frac{\Tobs T^{\text{max}}_{ic}}{T^{\text{SFT}}_i},
\end{equation}
where we show explicity that the SFT duration depends on the frequency
band $i$ while the coherence time depends on the frequency band $i$
and orbital parameter space cell $c$.  Note that this is an
overestimate of the absolute number of pairs, because we computed the
$\Tobs$ using the start and end times of the two parts of O3 rather
than an actual set of data segments reflecting the true duty cycle.
In addition to the total computing cost
$\sum_i\sum_c N^{\text{pair}}_{ic} N^{\text{tmplt}}_{ic}$ for each
lattice, we also plot the lattice points projected onto the
$\TascNow,\Porb$ or $\TascNow,\PorbShear$ plane, limiting attention
for the plots to the shaded cells in \figref{f:Ta_regions}.

\begin{figure}[t]
  \begin{center}
    \includegraphics[width=0.47\columnwidth]{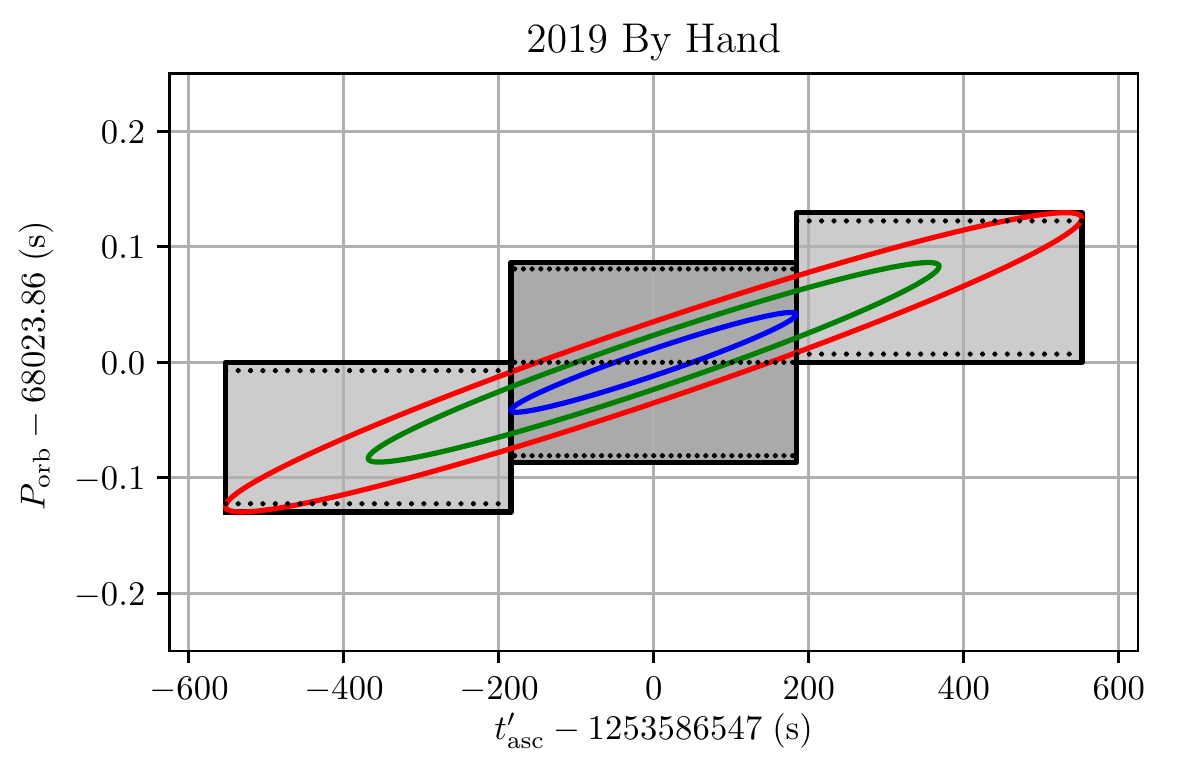}
    \includegraphics[width=0.47\columnwidth]{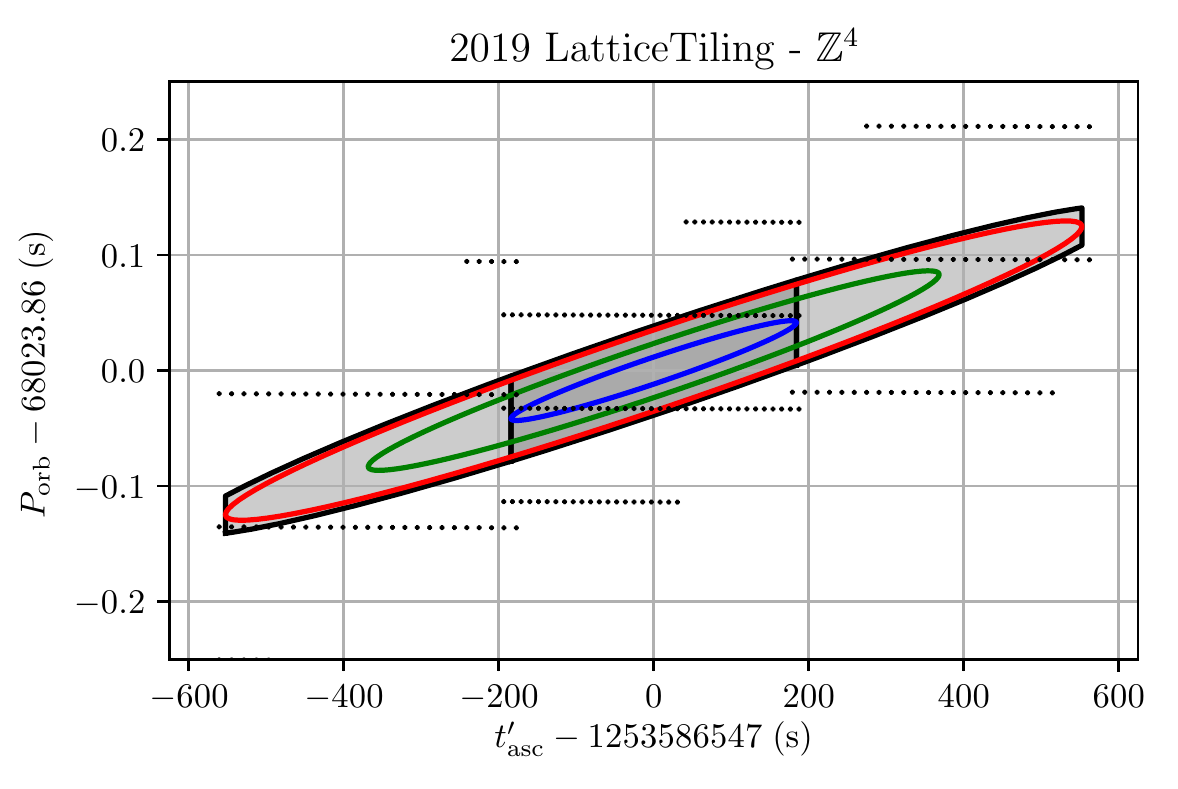}
  \end{center}
  \caption{Left: Lattice setups that most closely resemble what
    was done in previous CrossCorr searches, but with the inclusion of
    ``chopped'' search regions in  $\TascNow,\Porb$. The lattice points
    are placed by computing the spacing in each direction given the mismatch
    and the metric (\secref{s:byhand}). Points are then placed to cover the
    uncertainty ellipses in each of the three rectangular search regions,
    where the darker region bounded by one-sigma in $\TascNow$ represents the
    region where we are most likely to find a signal if it is present.The
    total number of templates for this setup is {\NumChopHand}. Right:
    Implementing the elliptical boundary function and using {\LatticeTiling}
    to place a cubic lattice changes where the points are placed. The total
    number of templates for this setup is {\NumCubeEllip}.}
  \label{f:9H9Z}
\end{figure}

\Figref{f:9H9Z} shows two implementations of cubic ($\mathbb{Z}^4$)
lattices, both using the original by hand method described in
\secref{s:byhand} and using the {\LatticeTiling} module.  The main
difference between the two methods is in how they handle the
boundaries of the elliptical search region.  The by-hand method
uses the chopped regions illustrated in the right panel of
\figref{f:TP_regions}, while the {\LatticeTiling} method uses the
elliptical boundaries of \figref{f:TP_regions_ellip}.  Note that while
{\LatticeTiling} uses a smaller region of parameter space, it actually
requires more templates (a total over the whole parameter space of
{\NumCubeEllip} versus {\NumChopHand} for the by-hand method) because
of its conservative approach to covering the boundaries.  Ordinarily
this would be a small effect, but since only two or three templates
are required in the $\Porb$ direction, it is significant in this case,
which motivates the special handing of the $\Porb$ coordinate which
follows.

\begin{figure}[t]
  \begin{center}
    \includegraphics[width=0.47\columnwidth]{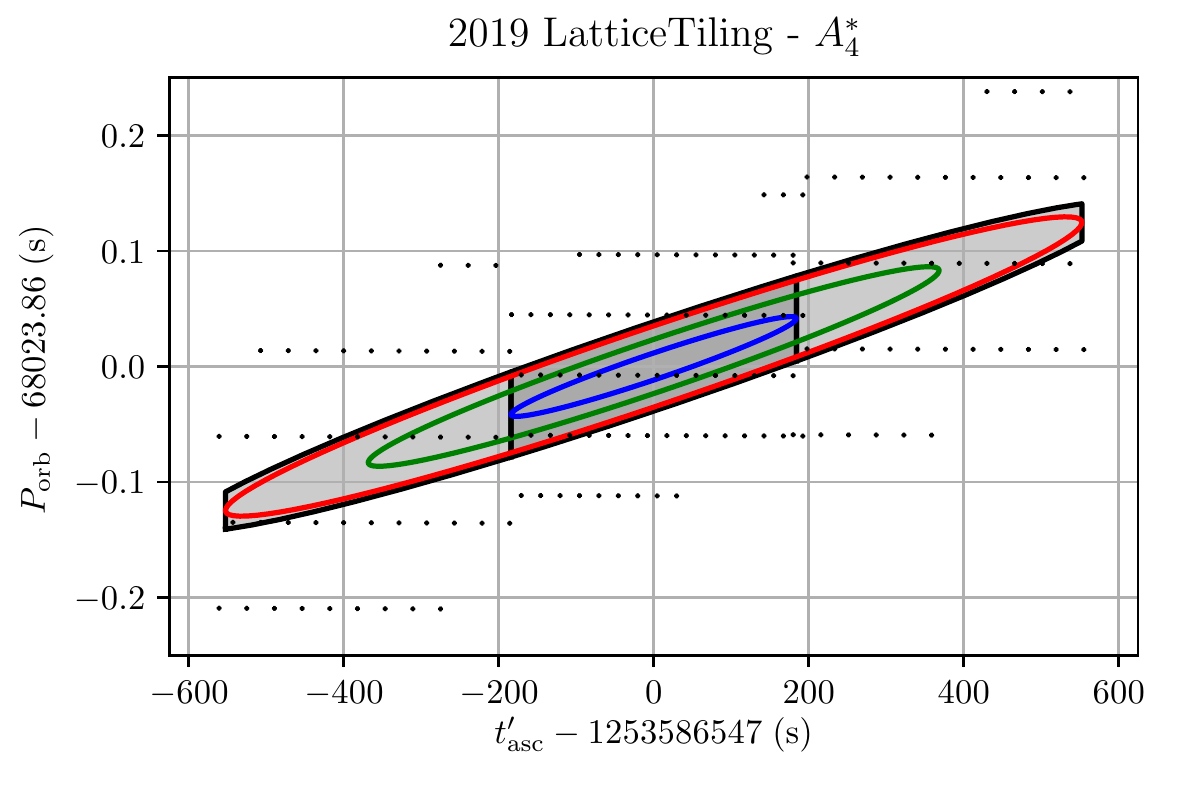}
    \includegraphics[width=0.47\columnwidth]{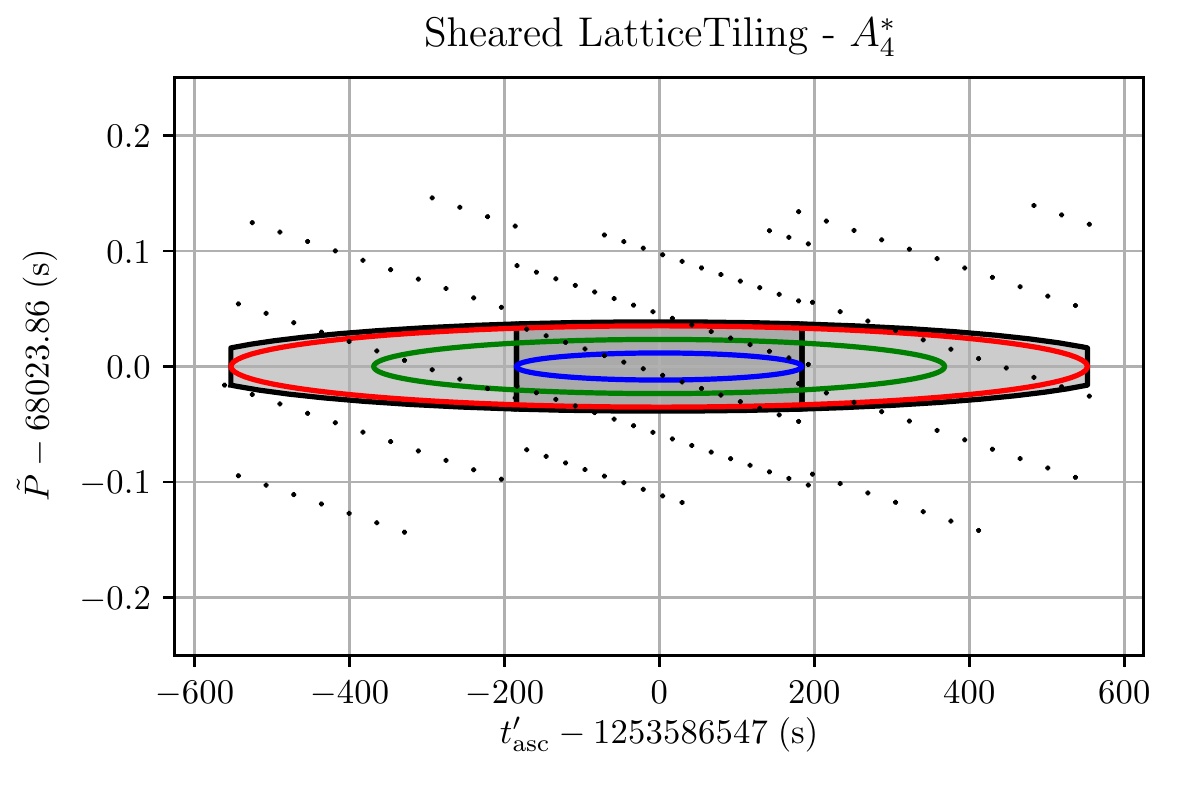}
  \end{center}
  \caption{Left: Implementing the elliptical boundary function
    discussed in \secref{ss:stdcoords} requires the use of
    {\LatticeTiling} \cite{Wette2014_Lattice} for template
    placement. This setup shows a use of an $A_4^*$ lattice with the
    elliptical boundary function and lattice template points placed by
    {\LatticeTiling}. Note that the density of templates is
    increased in the one-sigma region.  The total number of templates
    used here (across four-dimensional parameter space) is
    {\NumPropEllip}.  Right: After performing the shearing
    transformation discussed in \secref{ss:sheared}, we use
    {\LatticeTiling} to place templates in
    $\TascNow, \PorbShear$ space. This figure shows an $A_4^*$ lattice
    over the sheared uncertainty ellipses using the elliptical
    boundary function. Note that the primary axes of the uncertainty
    ellipses are aligned with the coordinate axes in this
    area-preserving transformation, and that template density is again
    greater in the one-sigma region.  The total number of templates
    here is {\NumShearFour}.}
  \label{f:SA9A}
\end{figure}

\begin{figure}[t]
  \begin{center}
    \includegraphics[width=0.47\columnwidth]{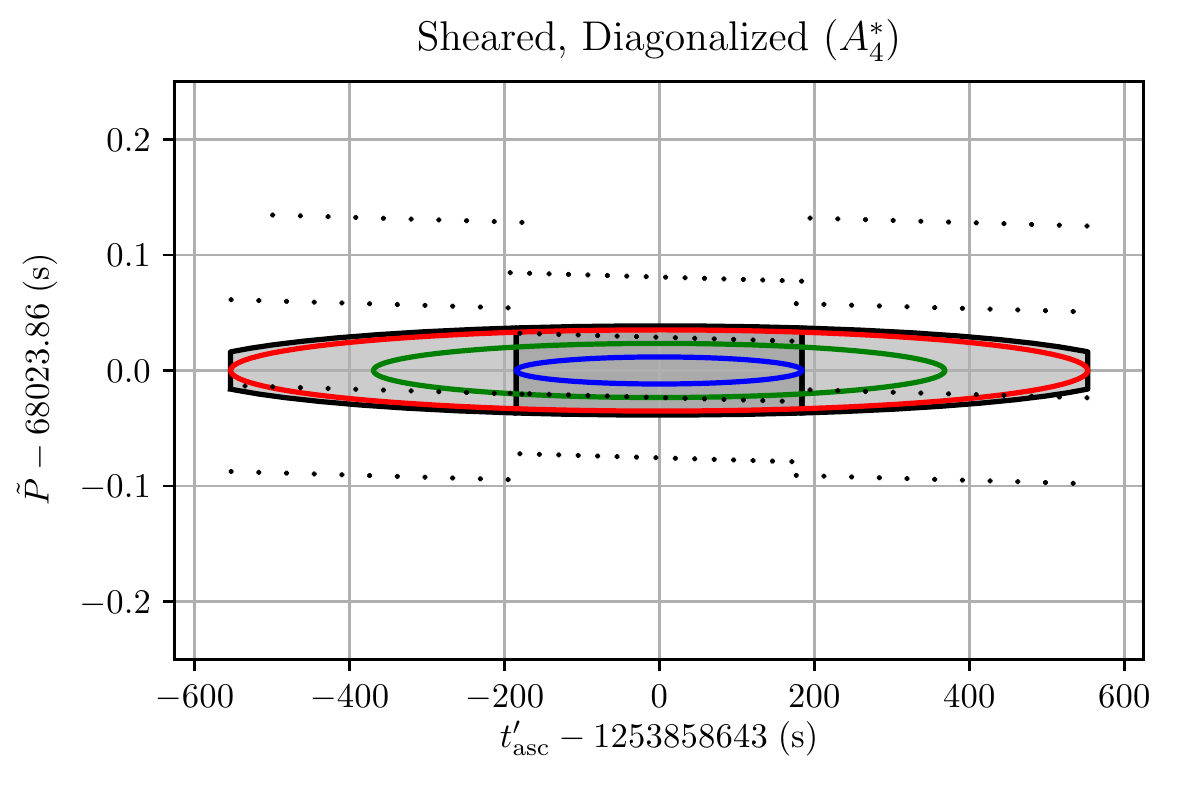}
    \includegraphics[width=0.47\columnwidth]{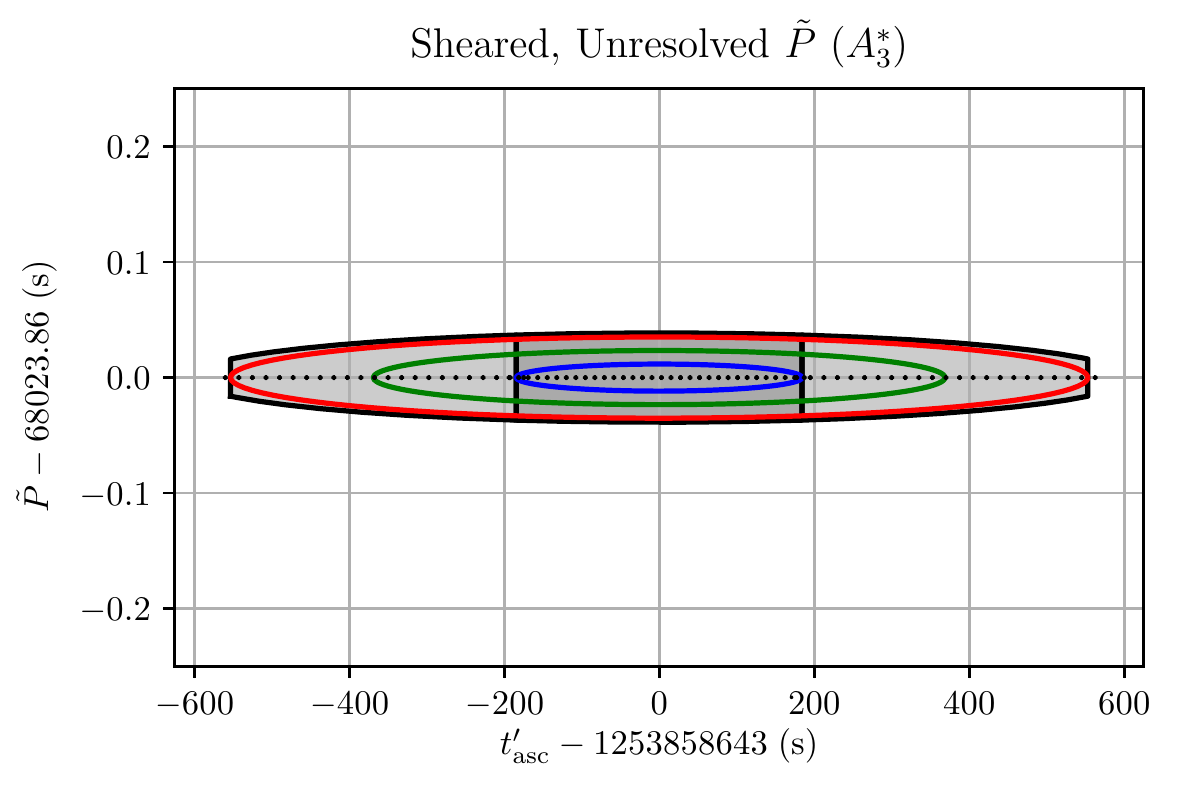}
  \end{center}
  \caption{Left: The number of orbits used to propagate $\TascThen$
    into 2019 coordinates was chosen based on what would diagonalize
    the standard/ 2019 metric. This is what introduced the slant in
    the template rows seen in \figref{f:SA9A}. Choosing a different
    $\norb$ eliminates the slant, to produce this figure, showing a
    lattice covering using an $A_4^*$ lattice, {\LatticeTiling}
    to place the templates, and the sheared coordinates with a
    diagonalized metric to align the uncertainty ellipses with the
    primary axes of the parameter space. The total number of templates
    is {\NumShearDiag}.  Right: Noticing that the spacing between
    template rows in $\TascNow, \PorbShear$ seemed to be larger than
    the cross-section of the uncertainty ellipse, we perform a
    calculation described in \secref{ss:sheared} to determine whether
    the orbital period needs to be resolved in the sheared
    coordinates. After finding that it does not, for our search, we
    fix $\PorbShear = P_0$, forcing {\LatticeTiling} to place a
    single row of lattice templates along the centerline of the
    sheared uncertainty ellipse.  Note that the template density is
    still greater in the one-sigma region. Here, the total number of
    templates is {\NumShearThree}, our best result for template count
    and an improvement from the original setup by a factor of about
    3.}
  \label{f:SPSD}
\end{figure}

If we change the lattice from $\mathbb{Z}^4$ to $A_4^*$, we obtain the
lattice shown in the left panel of \figref{f:SA9A}.  The use of a more
efficient lattice has reduced the total number of templates to
{\NumPropEllip}, but we can see from the figure that the templates
extend well beyond the boundaries of the search region.  In the right
panel, we construct the lattice in the sheared {\coord}s
$\TascNow,\PorbShear$ defined in \secref{ss:sheared}, which simplifies
the boundaries of the search region, but produces lattices with
comparable numbers of templates ({\NumShearFour} total).  In these
{\coord}s, the mismatch metric has a non-negligible off-diagonal
component $\gShearTP$, so the template lattice is constructed using a
basis which looks ``slanted'' in these {\coord}s.

We can make the
metric approximately diagonal, as described in \secref{ss:sheared}
by choosing a different value of $\norb$ derived from
\eqref{e:norb-sheared}; for the example considered, this means
changing $\norb$ from $\bestNorb$ to $\shearBestNorb$.  The resulting
lattice is shown in the left panel of \figref{f:SPSD}.  Note that the
total number of templates is comparable to the other $A_4^*$ lattices,
a total {\NumShearDiag} across the whole parameter space.  The fact
that all of the $A_4^*$ lattices have comparable numbers of templates
indicates that the {\LatticeTiling} module is behaving consistently,
even when the {\coord}s being used have metric
correlations or oddly-shaped boundaries.  However, it is clearly not
taking full advantage of the narrow range of plausible $\PorbShear$
values.  The underlying issue is that {\LatticeTiling}, by the nature
of its boundary-covering algorithm \cite{Wette2014_Lattice}, uses a
minimum of two templates in a {\coord} direction, even if a single
template would be sufficient to cover the space at the desired minimum
mismatch.

The change to $\TascNow,\PorbShear$ {\coord}s, in which both the prior
uncertainty and mismatch metric are approximately uncorrelated, allows
us to take advantage of the small prior uncertainty in $\PorbShear$.
If we limit attention to lattices with all their templates on the
hypersurface $\PorbShear=\BestPorb$, the mismatch between a signal
with parameters $\{\lambda^s_i\}$ and a template point $\{\lambda_i\}$
will be
\begin{equation}
  \mu = \gShearPP(\PorbShear^s-\BestPorb)^2 + 2\sum_{\alpha} g_{\alpha\PorbShear}
  (\lambda_{\alpha}^s-\lambda_{\alpha})(\PorbShear^s-\BestPorb)
  + \sum_{\alpha}\sum_{\beta}
  g_{\alpha\beta}(\lambda_{\alpha}^s-\lambda_{\alpha})
  (\lambda_{\alpha}^s-\lambda_{\alpha}),
\end{equation}
where $\{\lambda_{\alpha}\}=\{\TascNow,f_0,\ap\}$ are the other three
{\coord}s of the parameter space and $\gShearPP$ is the sheared metric
element for orbital period.  If we assume the metric is
approximately diagonal, this becomes
\begin{equation}
  \mu \approx \gShearPP(\PorbShear^s-\BestPorb)^2 + \mu^{\parallel},
\end{equation}
As shown in the Appendix, the general expression is
\begin{equation}
  \mu \approx \frac{(\PorbShear^s-\BestPorb)^2}{\ginvShearPP} + \mu^{\parallel}
\end{equation}
Since the prior uncertainty ellipse with $\chi^2\le k^2$ (see
\eqref{e:chisqShear} and \figref{f:TP_regions_ellip}) has
$(\PorbShear-\BestPorb)^2\le k^2\sigPorb^2$, we can obtain a lattice
with $\mu<\mumax$ everywhere if we construct a three-dimensional
lattice with
\begin{equation}
  \mumax^{\parallel} \le \mumax - \frac{k^2\sigPorb^2}{\ginvShearPP}
\end{equation}
A conservative approach is to allocate a mismatch of
$\frac{\mumax}{4}$ to the $\PorbShear$ direction and
$\frac{3\mumax}{4}$ to the other three directions.  Then we proceed as
follows:
\begin{itemize}
\item If $\frac{k^2\sigPorb^2}{\ginvShearPP}>\frac{\mumax}{4}$, we construct an
  $A_4^*$ lattice covering the full four-dimensional parameter space
  as usual.
\item If $\frac{k^2\sigPorb^2}{\ginvShearPP}\le\frac{\mumax}{4}$, we construct a
  three-dimensional $A_3^*$ lattice with maximum mismatch
  $\mumax^{\parallel}\frac{3\mumax}{4}$ and $\PorbShear=\BestPorb$ at
  all lattice points.  (In {\LatticeTiling} we accomplish this by
  setting the search region to have zero width in the $\PorbShear$
  direction.)
\end{itemize}
Following this approach produces the most efficient lattice, with
{\NumShearThree} total templates, illustrated in the right panel of
\figref{f:SPSD}.  A slightly more agressive approach would be to
``reallocate'' any unused mismatch if
$\frac{k^2\sigPorb^2}{\ginvShearPP}<\frac{\mumax}{4}$, and set to the
maxiumum mismatch of the $A_3^*$ lattice to
\begin{equation}
  \mumax^{\parallel} = \mumax - \frac{k^2\sigPorb^2}{\ginvShearPP}.
  \ ;
\end{equation}
This leads to a slightly smaller number of templates (\NumReallocThree).

\begin{table}[t]
  \centering
  \caption{Comparing Estimates of Raw Computing Cost: we display the chosen
    coordinates, the number of orbits needed to propogate $\TascNow$ to obtain
    a diagonal metric, the type of search region boundary used, and the type
    of lattice structure. Then, we show the number of templates required to
    cover all of parameter space using a given lattice and estimate the computing
    cost by multiplying the number of lattice templates by the number of SFT
    pairs.}
  \label{tab:results}
\begin{tabular}{ |c|c|c|c|c|c| }
      \hline
      {\Coord}s & $\norb$ & Boundary & Type
      & $\sum_{i,c} N^{\text{tmplt}}_{ic}$
      & $\sum_{i,c}N^{\text{pair}}_{ic} N^{\text{tmplt}}_{ic}$ \\
      \hline
      \hline
      $\TascNow,\Porb,\ap,f_0$ & $\bestNorb$ & Chopped & $\mathbb{Z}^4$
      & {\NumChopHand} & {\CostChopHand} \\
      \hline
      $\TascNow,\Porb,\ap,f_0$ & $\bestNorb$ & Elliptical & $\mathbb{Z}^4$
      & {\NumCubeEllip} & {\CostCubeEllip} \\
      \hline
      $\TascNow,\Porb,\ap,f_0$ & $\bestNorb$ & Elliptical & $A_4^*$
      & {\NumPropEllip} & {\CostPropEllip} \\
      \hline
      $\TascNow,\PorbShear,\ap,f_0$ & $\bestNorb$ & Elliptical & $A_4^*$
      & {\NumShearFour} & {\CostShearFour} \\
      \hline
      $\TascNow,\PorbShear,\ap,f_0$ & $\shearBestNorb$ & Elliptical & $A_4^*$
      & {\NumShearDiag} & {\CostShearDiag} \\
      \hline
      $\TascNow,\ap,f_0$; $\PorbShear=\BestPorb$ & $\shearBestNorb$ & Elliptical & $A_3^*$
      & {\NumShearThree} & {\CostShearThree} \\
      \hline
      \multicolumn{4}{|c|}{Same with reallocated mismatch}
                & {\NumReallocThree} & {\CostReallocThree} \\
      \hline
    \end{tabular}
\end{table}

The properties of the different lattices are summarized in
\tabref{tab:results}.  In addition to the total number of templates
$\sum_{i,c} N^{\text{tmplt}}_{ic}$ across all of the parameter space
cells, we also show the sum
$\sum_{i,c}N^{\text{pair}}_{ic} N^{\text{tmplt}}_{ic}$ which should
roughly scale with the computing cost.  Roughly speaking, replacing
the by-hand cubic lattice with an $\Ans$ lattice reduces to overall
computing cost by a factor of 2, while enforcing
unresolved $\PorbShear$ when possible reduces the cost by a further
factor of 1.5, for an overall improvement of a factor of
3 resulting from the enhancements described in this paper.

\section{Conclusions}
\label{s:conclusions}

In this paper we have discussed changes to the lattice used in the
template-based cross-correlation search for continuous gravitational
waves from Scorpius X-1. We detailed the setup of our parameter space
and explained how previous searches used lattices in the same
parameter space. We then gave four major improvements to
improve the lattice setup, using fewer templates for a given computing
cost. We first showed that there is a reduction in template count by
switching from a hypercubic lattice to an $\Ans$ lattice in
\secref{s:byhand}. Then, we defined an elliptical boundary function in
\secref{ss:stdcoords} to improve the shape of the search region in
$\TascNow$ and $\Porb$ to be more focused on the section of parameter
space within the prior ellipses. In \secref{ss:sheared} we defined an
area-preserving shearing transformation that aligned the axes of the
prior ellipses with the coordinate axes. This simplifies the task of
using {\LatticeTiling} to place a horizontal row of templates in
parameter space. Finally, we compared the cross-section of the prior
ellipses in $\TascNow$ and $\PorbShear$ to determine whether
$\PorbShear$ needed to be resolved, and determined that it did not in
\secref{ss:sheared}. This allowed us to use an $A_3^*$ lattice, and
reduced the original template count by a factor of $\sim 3$.
This reduction in template count allows the use of longer coherence
times at the same computing cost, enabling a more sensitive search.

\section*{Acknowledgments}

We wish to thank Chris Messenger, as well as the members of the LIGO
Scientific Collaboration and Virgo Collaboration continuous waves
group, for useful feedback.
KJW, JTW, and JKW were supported by NSF grant PHY-1806824.
KW was supported by the Australian Research Council Centre of Excellence for
Gravitational Wave Discovery (OzGrav) through project number CE170100004.
This paper has been assigned LIGO Document Number \dcc.

\section*{References}

\providecommand{\newblock}{}

\appendix

\section{Fixing the Sheared Period {\Coord}}

Consider how we handle the mismatch when $\PorbShear$ is
underresolved.  Let the search region be contained within the range
$\BestPorb - \Delta\PorbShear \le \PorbShear \le \BestPorb +
\Delta\PorbShear$ and let $\{\lambda_\alpha\}=\{f_0,a_p,\TascNow\}$ be
the other three search {\coord}s.  If we construct a template lattice
in $\{\lambda_\alpha\}$ with a maximum mismatch $\mumax^\parallel$, we
can ask what is the mismatch between a point in that lattice and a
point on the $\PorbShear$ boundary, with
$\PorbShear = \BestPorb + \Delta\PorbShear$.  If
$\Delta\lambda_{\alpha}$ is separation from a grid point, the total
mismatch will be
\begin{equation}
  \mu = \gShearPP(\Delta\PorbShear)^2 + 2\sum_{\alpha} g_{\alpha\PorbShear}
  (\Delta\lambda_{\alpha})(\Delta\PorbShear)
  + \underbrace{\sum_{\alpha}\sum_{\beta}
  g_{\alpha\beta}(\Delta\lambda_{\alpha})(\Delta\lambda_{\beta})}_{\mu^{\parallel}}.
\end{equation}
If the metric is approximately diagonal, this becomes
\begin{equation}
  \mu = \gShearPP(\Delta\PorbShear)^2 + \mu^{\parallel}.
\end{equation}
One conservative approach is to say that as long as
$\gShearPP(\Delta\PorbShear)^2 < \frac{\mumax}{4}$, we will set
$\PorbShear$ to $\BestPorb$ and define a template lattice in the other
three {\coord}s with $\mumax^{\parallel}=\frac{3}{4}\mumax$.  More
generally, we could choose
\begin{equation}
  \mumax^{\parallel} = \mumax - \gShearPP(\Delta\PorbShear)^2
\end{equation}
which will work as long as $\gShearPP(\Delta\PorbShear)^2 < \mumax$.

In general, though, the metric might not be diagonal, and in
particular $\gShearTP$ might be non-negligible.  To see how the
mismatch for a point on the $\PorbShear$ boundary changes, consider
the case of a two-dimensional lattice in $\TascNow$ and $\PorbShear$,
so that the mismatch is
\begin{equation}
  \mu = \gShearPP(\Delta\PorbShear)^2
  + 2 \gShearTP(\Delta\TascNow)(\Delta\PorbShear) + \gShearTT(\Delta\TascNow)^2.
\end{equation}
Suppose the spacing in the $\TascNow$ direction is
\begin{equation}
  \delta\TascNow = 2\sqrt{\frac{\mu^{\parallel}}{\gShearTT}}
  \ .
\end{equation}
Consider two adjacent lattice points separated by $\delta\TascNow$,
and a point in between them, which has $\Delta\TascNow=t>0$ from one
point and $\Delta\TascNow=t-\delta\TascNow<0$ from the second one.  A
point with this $\TascNow$ value and
$\PorbShear = \BestPorb + \Delta\PorbShear$ will have the maximum
possible mismatch if it is the same mismatch distance away from the
two nearest grid points.  This means we're looking for the $t$ which
solves
\begin{equation}
  \gShearPP(\Delta\PorbShear)^2
  + 2 \gShearTP(\Delta\PorbShear) t + \gShearTT t^2
  = \gShearPP(\Delta\PorbShear)^2
  + 2 \gShearTP(\Delta\PorbShear) (t-\delta\TascNow)
  + \gShearTT (t-\delta\TascNow)^2.
\end{equation}
A bit of cancellation gives us
\begin{equation}
  0
  = - 2 \gShearTP(\Delta\PorbShear)(\delta\TascNow)
  - 2 \gShearTT (\delta\TascNow) t + \gShearTT (\delta\TascNow)^2,
\end{equation}
or
\begin{equation}
  t =  \frac{\delta\TascNow}{2}
  - \frac{\gShearTP}{\gShearTT}(\Delta\PorbShear)
  \ .
\end{equation}
As a quick sanity check, we see that this reduces to
$t = \frac{\delta\TascNow}{2}$ when $\gShearTP=0$, as we expect.
Plugging this back into the mismatch equation gives
\begin{equation}
  \begin{split}
    \mumax &= \gShearPP(\Delta\PorbShear)^2
    + 2 \gShearTP(\Delta\PorbShear)
    \left(
      \frac{\delta\TascNow}{2}
      - \frac{\gShearTP}{\gShearTT}(\Delta\PorbShear)
    \right)
    + \gShearTT
    \left(
      \frac{\delta\TascNow}{2}
      - \frac{\gShearTP}{\gShearTT}(\Delta\PorbShear)
    \right)^2
    \\
    &=
    \gShearTT
    \left(
      \frac{\delta\TascNow}{2}
    \right)^2
    +
    \frac{\gShearPP\gShearTT-\gShearTP^2}{\gShearTT}
    (\Delta\PorbShear)^2
    = \mumax^{\parallel}
    + \frac{(\Delta\PorbShear)^2}{\ginvShearPP}
  \end{split}
\end{equation}
where $\tilde{g}^{ij}$ is the inverse matrix to $\tilde{g}_{ij}$.

\end{document}